\documentclass[aps,prb,twocolumn,notitlepage,showpacs,floatfix,superscriptaddress]{revtex4}
\usepackage{bm}
\usepackage{amsmath,amssymb,graphicx}
\usepackage[colorlinks=true,citecolor=blue,urlcolor=blue,linkcolor=blue]{hyperref}
\usepackage{braket,titlesec,natbib}
\usepackage{comment}
\usepackage{float}
\newcommand{\beq}{\begin{equation}}
\newcommand{\eeq}{\end{equation}}
\newcommand{\bea}{\begin{eqnarray}}
\newcommand{\eea}{\end{eqnarray}}
\newcommand{\bes}{\begin{subequations}}
\newcommand{\ees}{\end{subequations}}
\newcommand{\ve}{\varepsilon}

\newcommand{\bk}{{\bf k}}
\newcommand{\bp}{{\bf p}}

\newcommand{\nn}{\nonumber}
\newcommand{\bwt}{\begin{widetext}}
\newcommand{\ewt}{\end{widetext}}

\newcommand{\bse}{\begin{subequations}}
\newcommand{\ese}{\end{subequations}}

\begin{document}

\title{Effective lattice model for collective modes in a Fermi liquid with spin-orbit coupling}

\author{Abhishek Kumar and Dmitrii L. Maslov}
\affiliation{Department of Physics, University of Florida, Gainesville, Florida, 32611}

\date{\today}
\begin{abstract}
A Fermi-liquid (FL) with spin-orbit coupling (SOC) supports a special type of collective modes--chiral spin waves--which are oscillations of magnetization even in the absence of the external magnetic field. We study the chiral spin waves of a two-dimensional  FL in the presence of both the Rashba and Dresselhaus types of SOC and also  subject to the in-plane magnetic field. We map the system of coupled kinetic equations for the angular harmonics of the occupation number onto an effective one-dimensional tight-binding model, in which the lattice sites correspond to angular-momentum channels. Linear-in-momentum SOC ensures that the effective tight-binding model has only nearest-neighbor hopping on a bipartite lattice. In this language, the continuum of spin-flip particle-hole excitations becomes a conduction band of the lattice model, whereas electron-electron interaction, parameterized by the harmonics of the Landau function, is mapped onto lattice defects of both on-site and bond type. Collective modes correspond to bound states formed by such defects. All the features of the collective-mode spectrum receive natural explanation in the lattice picture as resulting from the competition between on-site and bond defects.  
\end{abstract}

\maketitle
\section{Introduction}
 Spin-orbit coupling (SOC) allows  an external electric field of either \textit{dc} current or electromagnetic wave to act directly on electron spins. This phenomenon, known as Electron Dipole Spin Resonance (EDSR), \cite{rashba:1991,wilamowski:2002,rashbaefros2003, efros2006, duckheim2006,schulte:2005,wilamowski2007,wilamowski:2008}  enables one to detect electron spins in low-dimensional structures with smaller number of electrons.\cite{ESR:book} In semiconductor heterostructures, the most relevant types of SOC are Rashba\cite{rashba1960,bychkov1984} and Dresselhaus\cite{dresselhaus} mechanisms arising due to the lack of inversion symmetry, which is broken  either by an interface between two dissimilar materials or by lattice structure, correspondingly. 
  
In the single-particle picture of  EDSR,\cite{wilamowski:2002,rashbaefros2003, efros2006, duckheim2006}
SOC provides a link between electron spins and the electric field but does not affecting the resonance frequency itself, which is still given 
by the Larmor frequency determined by the applied magnetic field, as in conventional Electron Spin Resonance (ESR).
It has recently been realized, however, that electron-electron interaction modifies this picture significantly. In the absence of SOC, 
the Larmor frequency is the $q=0$ end point of the Silin-Leggett collective mode\cite{silin1958,leggett1970,statphys,baym} in a partially polarized Fermi liquid (FL). The Kohn\cite{Kohn:1961,yafet:1963} theorem states that this frequency is protected from renormalization by electron-electron interaction. In the presence of SOC, Kohn theorem is not applicable and a number of new phenomena arises. 
First of all, a FL with SOC supports a new type of collective modes--chiral spin waves--even in the absence of the magnetic field.\cite{shekhter2005,ashrafi2012,zhang2013,maiti2014,maiti2015} There are three such waves which corresponds to waves of magnetization linearly polarized in three perpendicular directions. If the magnetic field is applied, the structure of the collective-mode spectrum becomes fairly complex.\cite{maiti2016} There is a critical value of the field at which the Zeeman energy is equal to level splitting due to SOC. At this point, the Fermi surfaces of spin-split states become degenerate and spin-flip excitations cost no energy. For fields weaker than the critical one, there are still three collective modes which disperse down with the field. For fields stronger than the critical
 one, there is only one collective mode which becomes eventually the Silin-Leggett mode in the strong-field limit.
 
It is worth pointing out that chiral spin waves (in the strong-field regime) have been observed in a series of recent Raman experiments on magnetically doped CdTe quantum wells.\cite{perez:2013,perez:2015,perez:2016} The theoretical interpretation
of these experiments has been provided in Refs.~\onlinecite{karimi:2016,maiti:2017}.

The main goal of this paper is to provide a transparent physical interpretation for the complex behavior of the spin chiral waves as a function of the magnetic field. We start off with an observation that there are two systems which, albeit being completely different from the physical point-of-view, have nevertheless very similar excitation spectra. The first system is a  familiar tight-binding model with defects (or its continuum limit), whose spectrum consists of the band and discrete levels of bound states located outside the band.  The second system is a FL whose spectrum consists of the continuum of particle-hole excitations and collective mode of any type (plasmon, zero sound, spin waves, chiral spin waves, etc.) located outside the continuum. One cannot help but asking if there is a unifying mathematical description for both these systems. 

We answer this question affirmatively by considering a particular type of the FL, relevant in the context of EDSR and Raman experiments mentioned above, i.e.,   a two-dimensional (2D)  FL with both Rashba and Dresselhaus types of SOC and subject to the in-plane magnetic field. We show that the kinetic equation for such a FL can be mapped onto an effective one-dimensional (1D) tight-binding model with both on-site and bond defects. In this mapping, the conduction band of the lattice model plays the role of the continuum of spin-flip particle-hole excitations, whereas bound states corresponds to the collective modes. Furthermore, the spin part of the non-equilibrium occupation with angular momentum $m$ becomes the Bloch wavefunction localized on site $m$ of the 1D lattice, the Rashba splitting plays the role of the on-site energy in an ideal lattice, whereas Zeeman and Dresselhaus splittings play the role of hopping amplitudes between the nearest and next-to-nearest neighbors, correspondingly. Finally, angular harmonics of the Landau function, which parameterizes the interaction in a FL, play the role of local defects: the $m^{\text{th}}$ harmonic of the Landau function ``damages'' sites $m$, $m\pm 1$, and $m\pm 2$, as well as the adjacent bonds. We show that all features of the collective-mode spectrum, including the merging of some of its branches with the continuum, receives a natural explanation with the lattice model as a result of the competition between on-site and bond defects. In addition, we also derived analytic results for collective modes in a FL with Rashba SOC and in the presence of the magnetic field for a model ($s$-wave) form of the Landau function and computed the spectra numerically for a number of more general models of the interaction.

The rest of the paper is organized as follows. In Sec.~\ref{sec:model}, we introduce the model and discuss the FL kinetic equation.
In Sec.~\ref{sec:mapping}, we discuss the general strategy of mapping onto an effective lattice model. In Sec.~\ref{sec:RSOC}, we discuss the exactly soluble case of a FL with Rashba SOC\cite{shekhter2005} from the effective-lattice point-of view. In Sec.~\ref{sec:Rashba+B}, we consider a FL with Rashba SOC subject to the in-plane magnetic field; both in the $s$-wave approximation for the Landau function (Sec.~\ref{sec:s-wave}) and for a general case (\ref{sec:beyond_s}). In Sec.~\ref{sec:interp}, we provide physical interpretation of the collective mode spectrum within the effective lattice model. In Sec.~\ref{sec:R+D}, we consider a FL with both Rashba and Dresselhaus types of SOC. Our conclusions are given in Sec.~\ref{sec:conc}. Some computational details are delegated to Appendices \ref{TB}-\ref{app:s_p_wave}.

\section{Model and formalism}
\label{sec:model}
\begin{figure}[htbp]
\centering
\includegraphics[scale=0.30]{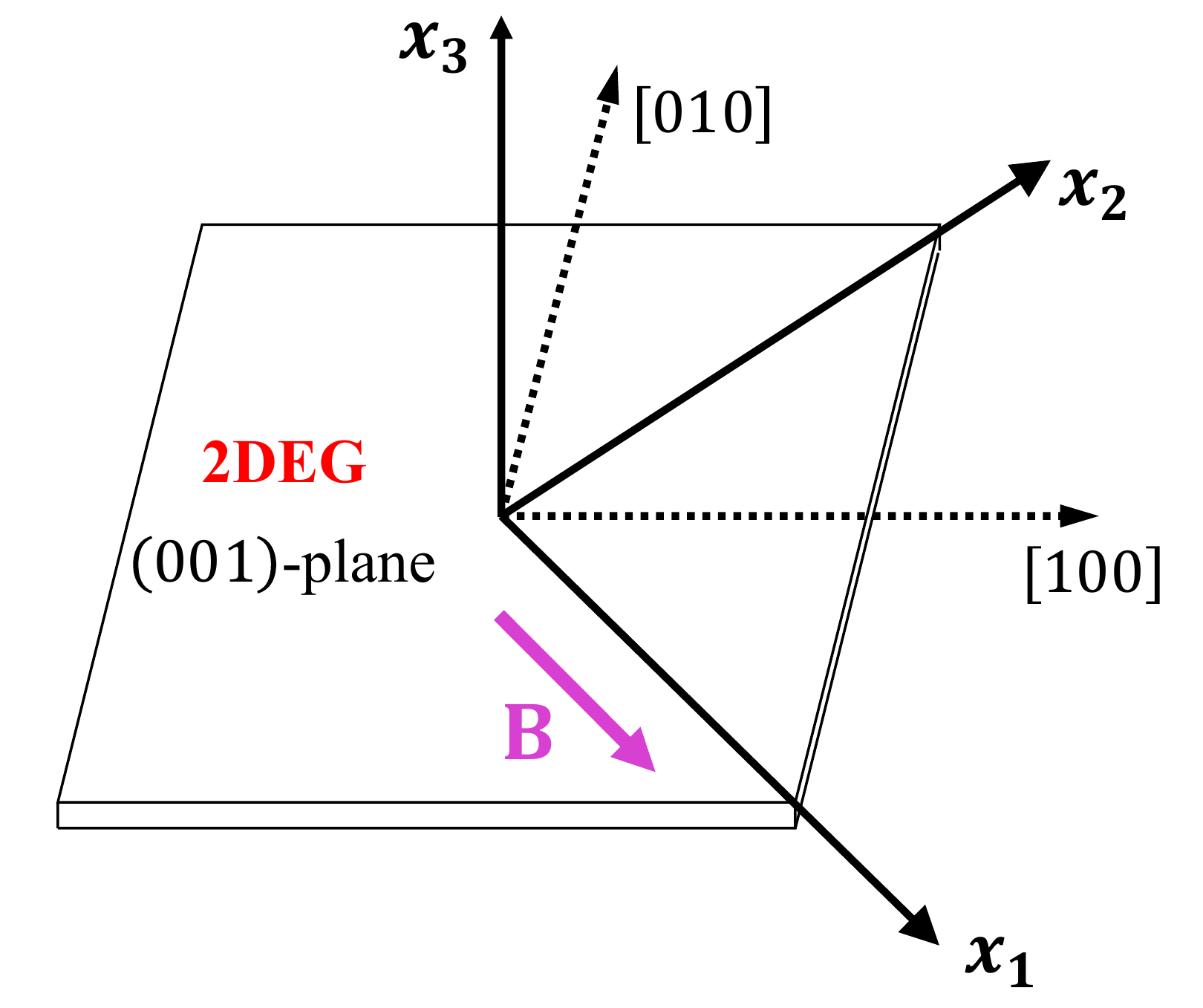}
\caption{\label{setup} Geometry of the system.}
\end{figure}

We consider a 2D  FL  in the presence of SOC of both the Rashba\cite{rashba1960, bychkov1984} and Dresselhaus\cite{dresselhaus} types, and subject to the in-plane static magnetic field. For a $(001)$ quantum well, the single-particle part of the Hamiltonian reads
\bea
\label{free hamiltonian}
\hat{H}_0&=& \frac{\textbf{p}^2}{2m_b}\hat{\sigma}_0 {\,} +\alpha\left(\hat{\sigma}_1p_2-\hat{\sigma}_2p_1\right)
+ \beta\left(\hat{\sigma}_1p_2+\hat{\sigma}_2p_1\right)\nn\\
&&
+
\frac{g\mu_B}{2}\hat{\sigma}_1B,
\eea
where $m_b$ is the band mass, $\hat{\sigma}_{1\dots 3}$ are the Pauli matrices, $\hat{\sigma}_0$ is the 2$\times$2 identity matrix, $\alpha$
($\beta$) 
is the Rashba
(Dresselhaus) 
coupling constant,  $\mu_B$ is the Bohr magneton, $g$ is the Land{\'e} factor, and $B$ is  the magnetic field. 
Furthermore, indices $1\dots 3$ label the axes of a Cartesian system with the $x_1$ and $x_2$ axes chosen to be along the $[1{\bar 1}0]$ and $[110]$ directions, respectively, see Fig. $\ref{setup}$. The magnetic
field is chosen to be along the high-symmetry axis ($x_1$). If the Dresselhaus term is absent ($\beta=0$), the system is invariant with respect to rotations about the $x_3$-axis. In this case, the direction of ${\bf B}$ is arbitrary, and the $x_1$-axis can be chosen along this direction.

We assume that both SOC and magnetic field are weak, in a sense that the corresponding energy scales are small compared to the Fermi energy.
In this case, one can neglect the effect of these perturbations on the Landau interaction function,\cite{shekhter2005, ashrafi:2013} which retains its  \textit{SU}(2)-invariant form: 
\bea\nu_F f_{\alpha\beta,\gamma\delta}(\theta_{\textbf{pp}'})=F^s (\theta_{\textbf{pp}'}) \delta_{\alpha\gamma}\delta_{\beta\delta}+F^a (\theta_{\textbf{pp}'})\hat{\bm\sigma}_{\alpha\gamma}\cdot\hat{\bm\sigma}_{\beta\delta},\nn\\
\eea
where $\theta_{\bp\bp'}=\theta_\bp-\theta_{\bp'}$, $\theta_\bk$ is the azimuthal angle of vector $\bk$,  and $\nu_F = m^*/\pi=m_b(1+F^s_1)/\pi$ is the renormalized density of states. Only the spin-dependent part of the Landau interaction function will be important for what follows. Note that crystalline anisotropy enters only via the Dresselhaus term in Hamiltonian (\ref{free hamiltonian}), whereas the underlying FL is considered as rotationally-invariant.

Both SOC and magnetic field will be treated as weak external perturbations imposed on an   \textit{SU}(2)-invariant FL. The self-consistent equation for the variation of the quasiparticle energy reads
\bea
\label{self-consistent}
\delta\hat{\varepsilon}\left(\textbf{p}\right)=\delta\hat{\varepsilon}_s\left(\textbf{p}\right)+\text{Tr}'\int \frac{d^{2}p'}{(2\pi)^2}\hat{f}_{\textbf{pp}'}\delta\hat{n}\left(\textbf{p}'\right),
\eea
where $'$ indicates the spin state of the quasiparticle with momentum \textbf{p}$'$, $\delta \hat n (\textbf{p})$  is the variation of the occupation number, and 
\bea
\label{de}
\delta\hat{\varepsilon}_s(\textbf{p})&=&\frac{1}{2}\Delta_R (e_2\hat{\sigma}_1 - e_1\hat{\sigma}_2)
+ \frac{1}{2}\Delta_D (e_2\hat{\sigma}_1 + e_1\hat{\sigma}_2)\nn\\
&& 
+ \frac{1}{2}\Delta_Z\hat{\sigma}_1
\eea 
is the variation of the quasiparticle energy due to SOC of both types  and external magnetic field. Here $\Delta_R=2\alpha p_F$, $\Delta_D=2\beta p_F$, and $\Delta_Z=g\mu_B B$ are the spin-orbit and Zeeman energy splittings, respectively, and $\textbf{e}=\textbf{p}/p$. (We choose $\Delta_Z$ to be non-negative, whereas $\Delta_R$ and $\Delta_D$ can be of either sign.) In equilibrium, Eq. $(\ref{self-consistent})$ is solved by an Ansatz $\delta\hat\ve(\bp)=\delta
\hat\ve^*_s(\bp)$ and $\delta\hat n(\bp)=\partial_\ve n_0\delta\hat\ve_s^*(\bp)$, where $n_0$ is the Fermi function and $\delta
\hat\ve^*_s(\bp)$ differs from $\delta\hat\ve_s(\bp)$ in Eq.~(\ref{de})  only in that the bare energy splittings are replaced by the renormalized ones. Expanding $F^a(\theta)$ in a series of 2D harmonics, 
\beq
F^a (\theta)=\Sigma_{m}F_{m}^{a}e^{im\theta},
\eeq
we obtain the renormalized energy splittings as\cite{chen1999,saraga:2005, shekhter2005}
\begin{equation}
\label{renorm}
\Delta_{R}^*=\frac{\Delta_R}{1+F_{1}^{a}} 
, {\,\,} \Delta_{D}^*=\frac{\Delta_D}{1+F_{1}^{a}},\Delta_{Z}^*=
\frac{\Delta_Z}{1+F_{0}^{a}}.
\end{equation}

To derive the equations of motion, it is convenient to introduce a set of rotated Pauli matrices \cite{shekhter2005}
\bea\tau^1(\bp)=-\sigma_3,\; \tau^2(\bp)=\textbf{e}\cdot\bm{\sigma},\; \text{and }\;\tau^3(\bp)=e_2\hat{\sigma}_1 - e_1\hat{\sigma}_2,\nn
\\
\label{tau}
\eea
 and decompose  
 $\delta\hat{n}\left(\textbf{p}\right)$ into the equilibrium and non-equilibrium parts as 
 \bea
 \delta\hat{n}\left(\textbf{p},t\right)=\partial_{\varepsilon}n_0[\delta\hat{\varepsilon}^*_s\left(\textbf{p}\right) + \textbf{u}(\theta_\bp,t)\cdot\hat{\boldsymbol{\tau}}],\label{dn}
 \eea
 In a spatially-uniform case and in the absence of the residual interaction between quasiparticles, the occupation number satisfies the quantum kinetic equation 
\begin{equation}
\label{kinetic-equation}
i\frac{\partial\delta\hat{n}\left(\textbf{p},t\right)}{\partial t}=[\delta\hat{\varepsilon}\left(\textbf{p},t\right),      \delta\hat{n}\left(\textbf{p},t\right)].
\end{equation}
This seemingly simple equation embodies complex quantum dynamics and, as will be shown below, allows for a transparent interpretation in terms of the effective lattice model. 
Substituting $\delta\hat{n}\left(\textbf{p}\right)$ from Eq.~(\ref{dn}) and $\delta\hat{\varepsilon}\left(\textbf{p}\right)$ from Eq.~(\ref{self-consistent}) into Eq. $\left(\ref{kinetic-equation}\right)$, and linearizing with respect to \textbf{u}, we obtain
\bea
\label{kinetic}
i
\partial_t\textbf{u}\cdot\hat{\boldsymbol{\tau}}
=
[\delta{\hat\varepsilon}^*_s, \,&\text{Tr}'\int
_{\theta_{\bp'}}
F^a\left(\theta_{\textbf{pp}'}\right)(\hat{\bm{\sigma}}\cdot\hat{\bm{\sigma}}')
(\textbf{u}'\cdot\hat{\bm{\tau}}')]\nn\\
&+[\delta{\hat\varepsilon}^*_s, \textbf{u}\cdot\hat{\bm{\tau}}],
\eea
where 
$\delta{\hat\varepsilon}^*_s\equiv \delta{\hat\varepsilon}^*_s(\bp)$, ${\bf u}\equiv{\bf u}(\theta_\bp,t)$, ${\bf u}'\equiv{\bf u}(\theta_{\bp'},t)$, $\hat{\bm{\tau}}'\equiv \hat{\bm{\tau}}'(\bp)$, and $\int_{\theta_\bk}\equiv\int (d\theta_\bk/2\pi)$.
Using the identities $\hat{\bm{\sigma}}\cdot\text{Tr}\left(\hat{\bm{\sigma}}\hat{\tau}_{i}'\right)=2\tau_{i}'$ and $[\hat{\tau}_{3}, \hat{\tau}_{2}']=-2i\hat{\tau}_{1}\cos\theta_{\textbf{pp}'}$,\cite{shekhter2005}   
we find that the components of vector  ${\bf u}\left(\theta_\textbf{p}, t\right)$ satisfy the following system of equations 
\begin{widetext}
\begin{subequations}
\begin{equation}
\label{u1}
\begin{split}
\frac{\partial u_1(\theta_\textbf{p})}{\partial t} = -& \Big[ \Delta_{R}^* + \Delta_{Z}^* \sin\theta_\textbf{p} - \Delta_{D}^* \cos2 \theta_\textbf{p} \Big] u_2(\theta_\textbf{p}) + \Big[ \Delta_{Z}^* \cos\theta_\textbf{p} + \Delta_{D}^* \sin2\theta_\textbf{p} \Big] u_3(\theta_\textbf{p}) \\
&- \int_{\theta_{\textbf{p}'}} F^{a}(\theta_{\textbf{pp}'}) \Big[ \Delta_{R}^* \cos(\theta_{\textbf{p},\bp'}
) + \Delta_{Z}^* \sin\theta_{\textbf{p}'} - \Delta_{D}^* \cos(\theta_{\textbf{p}} + \theta_{\textbf{p}'}) \Big] u_2(\theta_{\textbf{p}'}) \\
&+ \int_{\theta_{\textbf{p}'}} F^{a}(\theta_{\textbf{pp}'}) \Big[ \Delta_{R}^* \sin(\theta_{\textbf{p},\bp'})
 + \Delta_{Z}^* \cos\theta_{\textbf{p}'} + \Delta_{D}^*\sin(\theta_{\textbf{p}} + \theta_{\textbf{p}'}) \Big] u_3(\theta_{\textbf{p}'}),
\end{split}
\end{equation}
\begin{equation}
\label{u2}
\frac{\partial u_2(\theta_\textbf{p})}{\partial t} = \Big[ \Delta_{R}^* + \Delta_{Z}^* \sin\theta_\textbf{p} - \Delta_{D}^* \cos2\theta_\textbf{p} \Big] u_1(\theta_\textbf{p}) + \Big[ \Delta_{R}^* + \Delta_{Z}^*\sin\theta_\textbf{p} - \Delta_{D}^* \cos2\theta_\textbf{p} \Big] \int_{\theta_{\textbf{p}'}} F^{a}(\theta_{\textbf{pp}'})u_1(\theta_{\textbf{p}'}),
\end{equation}
\begin{equation}
\label{u3}
\frac{\partial u_3(\theta_\textbf{p})}{\partial t} = - \Big[ \Delta_{Z}^* \cos\theta_\textbf{p} + \Delta_{D}^* \sin2\theta_\textbf{p} \Big] u_1(\theta_\textbf{p}) - \Big[ \Delta_{Z}^* \cos\theta_\textbf{p} + \Delta_{D}^* \sin2\theta_\textbf{p} \Big] \int_{\theta_{\textbf{p}'}} F^{a}(\theta_{\textbf{pp}'}) u_1(\theta_{\textbf{p}'}).
\end{equation}
\end{subequations}
\end{widetext}
[From now on, the argument $t$ of ${\bf u}(\theta_\bk,t)$ will be suppressed.]
At the next step, we expand \textbf{u} and $F^a$ over a basis of angular harmonics $[\textbf{u}=\Sigma_{m} e^{im\theta}\textbf{u}^m \, \text{and} \, F^a(\theta)=\Sigma_{m} e^{im\theta} F_{m}^a]$ to obtain a system of finite-difference equations for $u^m_{i}$:
\begin{widetext}
\begin{subequations}
\bea
\frac{\partial u_{1}^m}{\partial t} &= &-\Delta_{R}^* u_{2}^m \bigg[ 1 +\frac{1}{2} \big(F_{m-1}^a + F_{m+1}^a \big) \bigg] - \frac{1}{2i}\Delta_{Z}^*\Big[u_{2}^{m-1} - u_{2}^{m+1} \Big] \Big[ 1+F_{m}^a \Big] + \frac{1}{2} \Delta_{D}^*\Big[ u_{2}^{m-2} \big(1+F_{m-1}^a \big) + u_{2}^{m+2} \big(1+F_{m+1}^a \big) \Big]\nn\\
&&+ \frac{1}{2i}\Delta_{R}^*u_{3}^m \Big[ F_{m-1}^a - F_{m+1}^a \Big] + \frac{1}{2}\Delta_{Z}^*\Big[ u_{3}^{m-1} + u_{3}^{m+1} \Big] \Big[ 1+F_{m}^a \Big] + \frac{1}{2i} \Delta_{D}^*\Big[ u_{3}^{m-2} \big(1+F_{m-1}^a \big) - u_{3}^{m+2} \big(1+F_{m+1}^a \big) \Big],\nn\\
\label{u1_m}\\
\frac{\partial u_{2}^m}{\partial t} &=& \Delta_{R}^*u_{1}^m \Big[ 1+F_{m}^a \Big] + \frac{1}{2i}\Delta_{Z}^* \Big[ u_{1}^{m-1} \big(1+F_{m-1}^a \big) - u_{1}^{m+1} \big(1+F_{m+1}^a \big) \Big] - \frac{1}{2}\Delta_{D}^* \Big[ u_{1}^{m-2} \big(1+F_{m-2}^a \big) + u_{1}^{m+2} \big(1+F_{m+2}^a \big) \Big],\nn\\
\label{u2_m}\\
\frac{\partial u_{3}^m}{\partial t} &=& -\frac{1}{2} \Delta_{Z}^* \Big[ u_{1}^{m-1} \big(1+F_{m-1}^a \big) +  u_{1}^{m+1} \big(1+F_{m+1}^a \big) \Big] -\frac{1}{2i} \Delta_{D}^* \Big[ u_{1}^{m-2} \big(1+F_{m-2}^a \big) - u_{1}^{m+2} \big(1+F_{m+2}^a \big) \Big].\nn\\
\label{u3_m}
\eea
\end{subequations}
\end{widetext}
Solution of Eqs.~(\ref{u1_m}-\ref{u3_m}) is the main subject of this paper.
An analytic solution is possible  in special cases, when only one of the three couplings--Rashba, Dresselhaus, and Zeeman--is present. \footnote{In the absence of both types of SOC, the system (\ref{u1_m}-\ref{u1_m}) does not look like a familiar set of macroscopic equations for the Silin-Leggett mode.\cite{silin1958,leggett1970,statphys,baym} The difference, however, is superficial and can be removed by rotating the basis of Pauli matrices.} If both types of SOC are absent, the problem reduces to the well-studied case of a  partially spin-polarized FL, which supports  the Silin-Leggett  collective mode.\cite{silin1958, leggett1970,statphys,baym}
If the magnetic field is absent and only type of SOC is present, the system (\ref{u1_m}-\ref{u3_m}) is also exactly soluble. This is the case of chiral spin modes--collective oscillations of magnetization in the absence of magnetic field--which have recently been studied in Refs.~\onlinecite{shekhter2005,ashrafi2012, maiti2014,zhang2013,maiti2015}.
In all other cases, Eqs.~(\ref{u1_m}-\ref{u3_m}) do not allow for an analytic solution. A numerical solution is, of course, possible, and will be discussed below.
However, an important insight into the nature of solutions is gained by noticing that the original problem can be mapped onto an effective lattice model.

\section{Quantum kinetic equation for a Fermi liquid\\ as an effective lattice model}
\label{sec:mapping}
Inspecting Eqs.~(\ref{u1_m}-\ref{u3_m}), we notice that they are similar to the Schroedinger equations for the tight-binding model on a 1D lattice. In this analogy, the angular momentum ($m$)
plays the role of the lattice site index, while $u^m_{1\dots 3}$ can be viewed as orbitals located on site $m$, with three orbitals per site. Furthermore, the Rashba terms (those proportional to $\Delta^*_R$) are ``local'', in a sense that the time derivative of the orbital on site $m$ is proportional to another orbital on the same site. Therefore, the Rashba terms play the role of on-site energies.  On the other hand, the Zeeman terms (those proportional to $\Delta_Z^*$) connect the time derivative of the orbital on site $m$ to those on sites $m\pm 1$. One can then view those terms as generating  ``hopping''  between the nearest  neighbors. In the same way, the Dresselhaus terms (those proportional to $\Delta_D^*$) generate hopping between next-to-nearest neighbors. 

It will be shown in more detail in Sec.~\ref{non_int} that, in the absence of the FL interaction, the system is mapped onto a standard 1D tight-binding model.  The band arising naturally in this model is nothing but the continuum of spin-flip particle-hole excitations. The width of the band is determined by the combinations of the three
energy scales: $\Delta_R$, $\Delta_D$, and $\Delta_Z$. 

Harmonics of the Landau function, $F^a_m$, enter system (\ref{u1_m}-\ref{u3_m}) in two ways: some of them affect the Rashba terms, responsible for on-site energy shifts, while others affect the Zeeman and
Dresselhaus terms, responsible for nearest- and next-to-nearest neighbors, correspondingly.  The effect of $F^a_m$ is {\em local}, i.e, harmonic $F^a_m$ affects only sites $m, m\pm 1$, and $m\pm 2$. Here comes another step in mapping: in the effective lattice model, the FL interaction plays the role of local ``defects'', which affect both on-site energies and adjacent bonds.

Defects of a 1D lattice produce bound states with energies outside the band. These bound states are nothing but the collective modes lying necessarily outside the continuum of particle-hole excitations. Therefore, studying a much simpler problem of bound states in 1D lattice, one can understand a much more complicated case of a FL with  SOC and in the presence of the magnetic field.

\begin{table*}
\caption{\label{table:I}Mapping of the Fermi-liquid kinetic equation onto an effective 1D tight-binding model.}
\centering
\begin{tabular}{|c|c|}
\hline
{\bf Fermi-liquid kinetic equation} & {\bf 1D tight-binding model} \\  
\hline
angular momentum $m$ & lattice site $m$\\
\hline
azimuthal angle of momentum $\bp$ ($\theta_\bp$) & quasimomentum\\
\hline
harmonic $m$ of occupation number $u^m_{1\dots 3}$ & orbitals on site $m$\\
\hline
Rashba spin-orbit coupling & on-site energy\\
\hline
Zeeman splitting & nearest-neighbor hopping\\
\hline
Dresselhaus spin-orbit coupling & next-to-nearest neighbor hopping\\
\hline
continuum of spin-flip particle-hole excitations & conduction band\\
\hline
harmonics of the Landau function & local defects\\
\hline
collective modes & bound states\\
\hline 
 \end{tabular}
\end{table*}

The harmonic content of $F^a(\theta)$ determines how many defects are created. For example, if the Landau function contains only the zeroth harmonic (the $s$-wave approximation), i.e,
\bea 
F^a(\theta)=F_0^a\;\text{or}\; F^a_m=\delta_{m,0} F^a_0,
\label{swave}
\eea only up to 5 central sites of the lattice ($m=0,\pm 1,\pm 2$) are replaced by defects. In the opposite case of a sharply peaked Landau function, i.e., when $F^a(\theta)\propto \delta(\theta)$ and $F^a_m$ does not depend on $m$, all impurities are identical and occupy all the sites. In this case, each site contains an identical defect. This means that the original lattice is simply replaced by a different one, and the bands of single-particle excitations and collective modes merge 
into a single band of the new lattice. A more realistic Landau function monotonically decreases with $\theta$, and thus $F^a_m$ decrease with $m$ as well. In the lattice language,
this is equivalent to having a {\em non-uniform ordered alloy}, in which stronger defects are located in the central region of the lattice, while weaker ones are located at the edges. 

The key elements of mapping between the two models are summarized in Table \ref{table:I}.  We found it less instructive to present the details of mapping in the most general case, when all the three couplings--Rashba, Dresselhaus, and Zeeman--are present. Instead, we will show how this mapping works for a number of special cases.

\section{Fermi liquid with Rashba spin-orbit coupling}
\label{sec:RSOC}
\begin{figure}[htbp]
\centering
\includegraphics[scale=0.26]{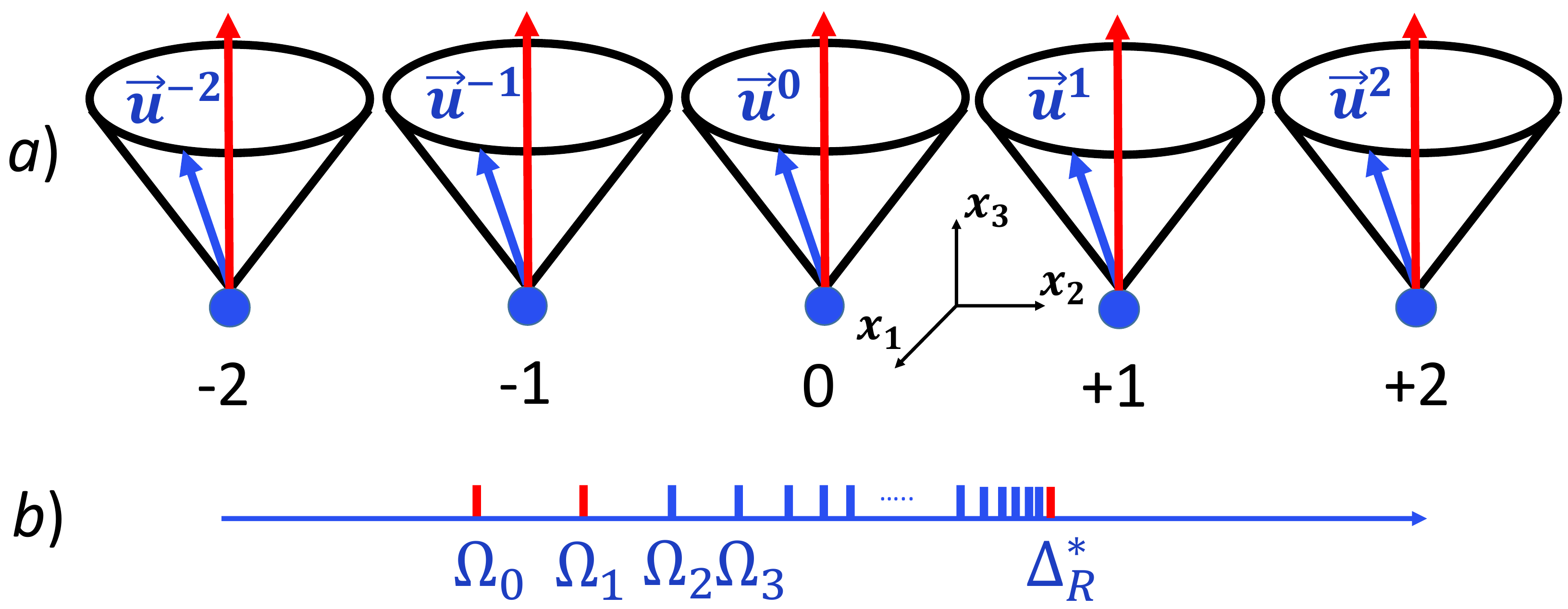}
\caption{\label{ESR_tightbinding} a) Effective lattice model for collective modes in a Fermi liquid with Rashba spin-orbit coupling. Lattice sites $0,\pm 1,\pm 2\dots$ correspond to the angular momenta  parametrizing the non-equilibrium part of the occupation number, ${\bf u}$ [Eq.~(\ref{dn})]. Each vector ${\bf u_m}$ represents a classical spin coupled to an effective spin-orbit magnetic field (vertical red arrows) via an anisotropic and site-dependent Land{\'e} factor. Spins precess independently of each other with site-dependent frequencies, which are the frequencies of the collective modes [Eq.~(\ref{freqR})]. b) The spectrum of the system consists of an infinite number of discrete levels, converging towards the continuum at $\Delta_{R}^*$.}
\end{figure}

To begin with, we consider the case when only Rashba SOC is present.
Although this case allows for an exact solution,\cite{shekhter2005} it is still beneficial to understand it within the effective lattice model.
With $\Delta_Z^*=\Delta_D^*=0$, Eqs.~(\ref{u1_m}-\ref{u3_m}) are reduced to
\bea
\partial_t u^m_1=-\gamma_1^m\frac{\Delta_R^*}{2} u_2^m;\;
\partial_t u_2^m=\gamma_2^m\frac{\Delta_R^*}{2} u_1^m;\;
\partial_t u_3^m=0,\nn\\
\eea
where $\gamma^m_1=2+ F_{m+1}^a+F_{m-1}^a$ and $\gamma^m_2=2(1+F_m^a)$.
In the absence of the Zeeman and Dresselhaus terms, there is no hopping between the sites of the effective lattice: the time evolution of $u_{1,2}^m$ is determined by $u^m_{2,1}$ on the same site.
In this case, one can think of vector ${\bf u}^m$ as a classical spin on site $m$. Spins do not interact with each other but are subject to a fictitious ``magnetic field" due to Rashba SOC, directed along the $x_3$ axis and of magnitude $\Delta_{R}^*/2$.  The effective Land{\'e} factor of these spins is anisotropic in the $(x_1,x_2)$ plane with components $\gamma^m_1$ and $\gamma^m_2$ given above. 
The lattice is also {\em non-uniform} because $\gamma_1^m$ and $\gamma_2^m$ depend on the lattice site.  Both anisotropy and site-dependence  of the $g$-factor arise from the FL interaction. With $u^m_3=\text{const}$,
the spins precess around the Rashba field,  see Fig.~\ref{ESR_tightbinding}a.

 A spin on site $m$ precesses with frequency
\bea
 \Omega_m=\Delta_{R}^*\sqrt{\left[1+\frac 12(F_{m+1}^a + F_{m-1}^a)\right](1+F_m^a)}.
 \label{freqR}
 \eea
These are the frequencies of the collective modes--chiral spin resonances.\cite{shekhter2005}  
The continuum of particle-hole excitation in the Rashba-only case is represented by a single frequency $\Omega=\Delta_R^*$. \footnote{If one accounts for ${\cal O}(\alpha^2)$ effects, the continuum acquires a finite width proportional to the Rashba energy, $m_b\alpha^2$.  However, as we mentioned before, such effects are neglected here.}
For any realistic interaction, the harmonics of $F^a(\theta)$ decrease with $m$; also, for a repulsive electron-electron interaction, $F_m^a < 0$ and thus $\Omega_m\leq \Delta^*_R$. Hence the discrete spectrum of the collective modes converges to the point of continuum, $\Delta_{R}^*$, from below, see Fig. $\ref{ESR_tightbinding}$b. 

The macroscopic magnetization is related to ${\bf u}(\theta_\bp)$ via 
\begin{equation}
{\bf M}=\frac{g\mu_B}{4}\nu_F\text{Tr}\int_{\theta}\hat{\bm{\sigma}}(\textbf{u}\cdot\hat{\bm{\tau}}),
\end{equation}
As is obvious from the form of the $\tau$-matrices [Eq.~(\ref{tau})], $\textbf{M}$ contains only the $m=0$ and $m=\pm1$ harmonics of $\textbf{u}$.
 In other words, external magnetic and electric fields couple only to the effective classical spins on sites $m=0,\pm 1$.
 These spins precess with frequencies
 \beq
 \Omega_0=\Delta_R^*\sqrt{\left(1+F_1^a\right)\left(1+F^a_0\right)}\label{om0}
 \eeq
  and 
  \beq
  \Omega_{+1}=\Omega_{-1}=\Delta_R^*\sqrt{\left[1+\frac 12 \left(F_0^a+F_2^a\right)\right]\left(1+F_1^a\right)}.\label{om1}\eeq
 These are the frequencies that should be observable by ESR in zero magnetic field. 

\section{Partially-polarized Fermi liqud\\with Rashba spin-orbit coupling}
\label{sec:Rashba+B}
\subsection{Non-interacting electrons}
\label{non_int}
\begin{figure*}
\centering
\includegraphics[scale=0.54]{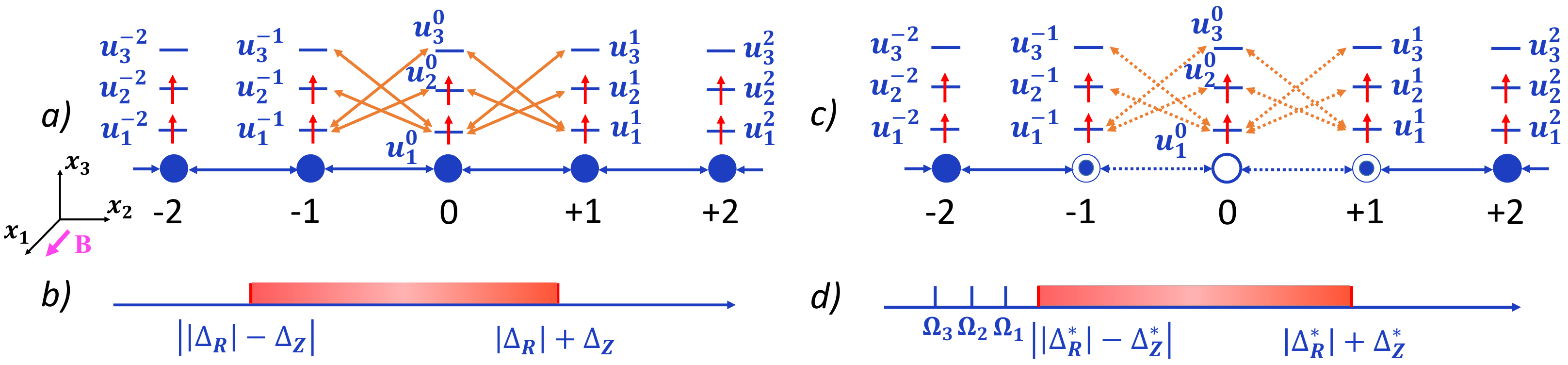}
\caption{\label{ESR_tightbinding_impurity} a) Three-orbital tight-binding model for a {\em non-interacting} electron gas with Rashba spin-orbit coupling and in the presence of the in-plane magnetic field ($B$). As in Fig.~\ref{ESR_tightbinding}, lattice sites $0,\pm 1,\pm 2\dots$ correspond to the angular momenta  parametrizing the non-equilibrium part of the occupation number, ${\bf u}$ [Eq.~(\ref{dn})] but now components $u_{1...3}^m$ play the role of on-site orbitals. The Rashba field (red vertical arrows) is coupled to orbitals $u_1^m$ and $u_2^m$, while $B$ leads to nearest-neighbor hopping with orbital-selective matrix elements. Allowed hoppings between sites -1, 0, and +1 are indicated by orange arrows, and similar for other sites. b) The spectrum consists of a finite-width band (shaded rectangular), which corresponds to the continuum of spin-flip particle-hole excitations. c) Same as in a) but for the FL case. The FL interaction, parametrized by the harmonics of the Landau function, $F_m^a$, create defects of both on-site and bond type.  If $F^a$ has only the $m=0$ harmonic, the defects (open and open-doted circles) are located at $m=0$ and $m=\pm 1$, as shown in the figure; higher harmonics of $F^a$ affect other sites. d) The spectrum consists of the continuum (shaded) and discrete bound states, which are the collective modes of the FL.}
\end{figure*}

Hopping between the sites of the effective lattice arises if at least two out of three couplings--Rashba, Dresselhaus, and Zeeman-- are present.  To understand the effect of hopping, we first consider the non-interacting case, when $F^a(\theta)=0$. To simplify the problem even further, we eliminate Dresselhaus SOC at first and restore it at the end of this section. With these simplifications, Eqs.~(\ref{u1_m}-\ref{u2_m})
are reduced to
\bea
\label{non-int}
\partial_{t}u_1^m &=& -\Delta_{R}u_2^m - \Delta_{Z}\bigg( \frac{u_2^{m-1} - u_2^{m+1}}{2i} - \frac{u_3^{m-1} + u_3^{m+1}}{2} \bigg),\nn \\ 
\partial_{t}u_2^m &=& \Delta_{R}u_1^m + \frac{1}{2i}\Delta_{Z}(u_1^{m-1} - u_1^{m+1}),\nn\\ 
\partial_{t}u_3^m &=& -\frac{1}{2}\Delta_{Z}(u_1^{m-1} + u_1^{m+1}).
\eea
One can interpret these equations as an effective tight-binding model with three orbitals,  $u^m_{1\dots 3}$, per site, see Fig.~\ref{ESR_tightbinding_impurity}a.  The Rashba spin-orbit field (shown by vertical red arrows) couples to the $u_1$ and $u_2$ orbitals but not  to the $u_3$ one. The Zeeman term gives rise to {\em orbital-selective} hopping between the nearest neighbors;  allowed hoppings are indicated by slanted orange arrows. For example, orbital $u_1$ is coupled to the orbitals $u_2$ and $u_3$, but there is no hopping between the orbitals $u_2$ and $u_3$. 

The tight-binding picture is simplified considerably by eliminating the orbitals $u_2^m$ and $u_3^m$ in favor of $u_1^m$.  Doing so, we arrive at a much simpler equation for 
temporal Fourier transform of $u_1^m$:
\beq
\Omega^2u_1^m=\left(\Delta_R^2+\Delta_Z^2\right)u_1^m+i\Delta_R\Delta_Z\left(u_1^{m+1}-u_1^{m-1}\right).\label{u1mB}
\eeq
Equation (\ref{u1mB}) is reduced to a standard tight-binding form 
by introducing the ``Bloch wavefunction'' 
\beq
\psi_m\equiv i^{-m} u_1^m,\label{vm}
\eeq
which satisfies\footnote{One might equally well view Eq.~(\ref{v1mB}) as describing a harmonic chain, but the relation between the coefficients of the diagonal and off-diagonal terms 
is not the same as for the standard case.  This can be fixed by connecting each ``atom'' by ``springs'' not only to its nearest neighbors but also
to its own site (by a spring with a different spring constant). However, we find the tight-binding interpretation of Eq.~(\ref{v1mB}) to be more straightforward.}
\beq
\Omega^2\psi_m=\left(\Delta_R^2+\Delta_Z^2\right)\psi_m-\Delta_R\Delta_Z\left(\psi_{m+1}+\psi_{m-1}\right).\label{v1mB}
\eeq
The eigenfrequency of Eq.~(\ref{u1mB}) 
\bea
\Omega(\theta_\bp)
&=&\left[\Delta_R^2+\Delta_Z^2-2\Delta_R\Delta_Z
\cos\theta_\bp\right]^{1/2}
\label{free}
\eea
disperses with $\theta_\bp\in(0,2\pi)$, which is a conjugate variable to $m$. Therefore, $\theta_\bp$ plays the role of ``quasimomentum" confined to the first Brillouin zone $(0,2\pi)$. 
At the same time, $\theta_\bp$ is nothing else but the azimuthal angle of $\bp$, so we worked our way back to original system (\ref{u1}-\ref{u3}),
which can now be viewed as written down in the ``momentum representation".

The minimum and maximum values of $\Omega(\theta_\bp)$
mark the edges of the ``band'', which corresponds to the continuum of spin-flip  excitations (shaded rectangular in Fig.~\ref{ESR_tightbinding_impurity}b).
The bandwidth is given by
  \beq\Omega_c
  =|\Delta_R|+\Delta_Z-||\Delta_R|-\Delta_Z|.\eeq
Likewise, with both SOC present but in the absence of the magnetic field, the band occupies an interval from $||\Delta_R|-|\Delta_D||$ to $|\Delta_R|+|\Delta_D|$.

Of course, one can obtain the same results in the momentum representation, i.e., directly from Eqs.~(\ref{u1}-\ref{u3}). Setting $F^a(\theta)=0$ and $\Delta_D=0$ in these equations, we obtain
\begin{equation}
\label{band dispersion}
\begin{split}
\partial_t u_1(\theta_\textbf{p})
&= - \left(\Delta_{R} + \Delta_{Z} \sin\theta_\textbf{p} 
\right) u_2(\theta_\textbf{p}) +
\Delta_{Z} \cos\theta_\textbf{p} 
u_3(\theta_\textbf{p}), \\
\partial_t u_2(\theta_\textbf{p})
&= \big(\Delta_{R} + \Delta_{Z} \sin\theta_\textbf{p}  
 \big) u_1(\theta_\textbf{p}), \\
\partial_tu_3(\theta_\textbf{p})
&= - 
\Delta_{Z} \cos\theta_\textbf{p} 
 u_1(\theta_\textbf{p}).
\end{split}
\end{equation}
Eliminating $u_2$ and $u_3$ from equations above, one obtains a second-order equation for $u_1$ which is an equivalent of Eq.~(\ref{u1mB}) in the momentum space. [The only difference between the two results is a $\pi/2$ shift of $\theta_\bp$ which is effected by transformation (\ref{vm}).]

The physical reason for the continuum to have a finite width is anisotropy of the electron spectrum in the presence of at least two couplings. At $q=0$, particle-hole excitations correspond to vertical transitions between spin-split subbands with frequencies  
\beq
\Omega_c=|\ve_{+}-\ve_{-}|,\label{exc}
\eeq 
where 
\bea
\ve_{\pm}&=&\frac{p^2}{2m_b}\pm\left[\left(\alpha^2+\beta^2\right)p^2+\frac{\Delta_Z^2}{4}\right.\nn\\
&&\left.+(\alpha+\beta) p_2\Delta_Z-2\alpha\beta \left(p_1^2-p_2^2\right)\right]^{1/2}
\label{epm}
\eea
are the eigenenergies of Hamiltonian (\ref{free hamiltonian}). As $\bp$ spans the FS, the eigenenergies vary between the minimum and maximum values. This variation determines
the width of the continuum. 
Setting $\beta=0$ and $p=p_F$ in Eq.~(\ref{epm}), we see that Eq.~(\ref{exc}) gives the same result as Eq.~(\ref{free}).

Restoring Dresselhaus SOC in the equations of motion does not lead to qualitative changes. With all the three couplings present, Eq.~(\ref{free}) for the eigenfrequency is replaced by
\bea
\Omega(\theta_\bp)
&=&\left[\Delta_R^2+\Delta_D^2+\Delta_Z^2+2(\Delta_R+\Delta_D)\Delta_Z\sin\theta_\bp\right.\nn\\
&&\left.-2\Delta_R\Delta_D\cos2\theta_\bp\right]^{1/2}.
\label{freeD}
\eea
As before, the maximum and minimum values of $\Omega(\theta_\bp)$ determine the bandwidth, but its explicit form is now more complicated and we refrain from presenting it.

In the special case when two out of the three couplings are absent, the spectrum becomes isotropic and the continuum shrinks to a single point. One case of this type was discussed in Sec.~\ref{sec:RSOC}.
\begin{figure}
\centering
\includegraphics[scale=0.40]{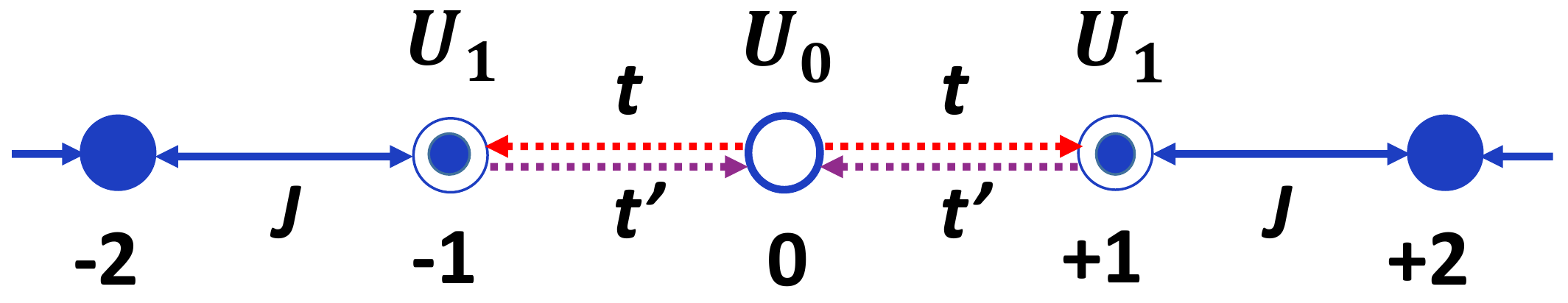}
\caption{\label{chiral} Single-orbital tight-binding model for a two dimensional FL  in the presence of Rashba spin-orbit coupling and in-plane magnetic field, and in the $s$-wave approximation for the Landau function [Eq.~(\ref{swave}.] Lattice $m=0$ and $\pm 1$ are defective. Defective bonds, connecting these sites are \textit{chiral}: the amplitude of hopping from $m=0$ to $m=\pm 1$ (dotted orange arrows) is not the same as from $m=\pm 1$ to $m=0$ (dotted green arrows).}
\end{figure}

\subsection{Interacting electrons}
We now turn to a FL with Rashba SOC and in the presence of the magnetic field. This case is described by Eqs.~(\ref{u1_m}-\ref{u2_m})
with $\Delta^*_D=0$. Pictorially, the equations of motion for $u_i^m$ are shown in Fig.~\ref{ESR_tightbinding_impurity}c. A harmonic $m$ of
of the Landau function changes both on-site energies on sites $m$ and $m\pm 1$ (shown by open and dotted-open circles for $m=0$) and the adjacent bonds (shown by dotted arrows). Both on-site and bond defects leads to formation of bound states with energies below the continuum, see Fig.~\ref{ESR_tightbinding_impurity}d.

To analyze this case quantitatively, we eliminate again  $u_2^m$ and $u_3^m$ from Eqs.~(\ref{u1_m}-\ref{u3_m}) in favor of $u_1^m$ and obtain an equation for the Bloch wavefunction $\psi_m$, defined
by Eq.~(\ref{vm}):
\begin{widetext}
\begin{equation}
\label{tight binding}
\begin{split}
\Omega^2\psi_m =& \left[ \Delta_{R}^{*2} \big( 1+F_m^a \big) \left( 1+\frac{F_{m+1}^a + F_{m-1}^a}{2} \right) + \Delta_{Z}^{*2} \big( 1+F_m^a \big)^2 \right] \psi_m \\
&-
\Delta_{R}^{*} \Delta_{Z}^{*}
\left[
\left( 1 + \frac{F_{m}^a + F_{m+1}^a}{2} \right) \left( 1 + F_{m+1}^a \right) 
\psi_{m+1} +
\left( 1 + \frac{F_{m-1}^a + F_m^a}{2} \right) \left(1 + F_{m-1}^a \right) 
 \psi_{m-1}\right] ,
\end{split}
\end{equation}
\end{widetext} 
Both the on-site and hopping terms are renormalized by the FL interaction.  The equation above  is the key one from which all the limiting cases can be derived, which is what we will be doing in the rest of this section. Before going into particular models for $F_m^a$ though, we make one general observation. An attractive impurity on a 1D lattice has at least one bound state below the band, while a repulsive one has at least one bound state above the band. 
If all $F^a_m<0$,
the on-site energies in Eq.~(\ref{tight binding}) are reduced compared to the non-interacting case [cf.~Eq.~(\ref{v1mB})]. This case corresponds to attractive impurities, with bound states (collective modes) below the band (continuum); it is vice versa for $F_m^a>0$, when the impurities are repulsive and the bounds states are above the band. On the other hand,  the number of bound states and their relative spacings will be specific for particular models, which we are now going to study.
\subsection{$s$-wave approximation for the Landau function}
\label{sec:s-wave}
 The first case is the $s$-wave approximation for the Landau function [Eq.~(\ref{swave})],
 when Eq.~(\ref{tight binding}) is reduced to
 \begin{widetext}
 \bea
 \Omega^2 \psi_m&=&\left(\Delta_{R}^{2} + \Delta_{Z}^{*2}\right)\psi_m +\delta_{m,0}F_0^a\left[\Delta_{R}^{2}+\Delta_{Z}^{*2}\left(2+F_0^a\right)\right]\psi_m
 +\left(\delta_{m,1}+\delta_{m,-1}\right)\frac{F_0^a}{2} \Delta_{R}^{2}\psi_m\nn\\
 &&-\Delta_R\Delta_Z^*\left\{\psi_{m+1}+\psi_{m-1}+\delta_{m,0}\frac{F^a_0}{2}\left(\psi_1+\psi_{-1}\right)+\left(\delta_{m,1}+\delta_{m,-1}\right)\frac{F^a_0}{2}\left(3+F_0^a\right)\psi_0\right\}.
 \label{tight binding s-wave}
\eea
\end{widetext}
We re-arranged the equation above in such a way that the first line of its RHS correspond to on-site energies while the second line correspond to hopping. [According to Eq.~(\ref{renorm}),  $\Delta_R$ is not renormalized in the $s$-wave approximation.]    Equations.~(\ref{tight binding s-wave}) is shown pictorially in Fig.~~\ref{chiral}. The first term in the first line of the RHS describes on-site energies of undistorted lattice, while  the remaining two terms account for energy shifts due to impurities at sites $m=0$ and $m\pm 1$ (open and dotted-open circles).  Next, the first term in the curly brackets describes hopping via regular bonds (solid arrows), while the remaining two terms describe  hopping via defective bonds (dashed arrows) which connect the
$m=0$ and $m=\pm 1$ sites. The bond defects are {\em chiral}: the  amplitude of hopping from $m=0$ to  $m=\pm 1$ is not the same as from $m=\pm 1$ to $m=0$, as indicated by one-handed arrows. This means that the effective Hamiltonian corresponding to the equations of motion (\ref{tight binding s-wave}) is non-Hermitian: bond defects are described by a non-Hermitian term
$J\Psi^\dagger_0\Psi_{\pm 1}+J'\Psi^\dagger_{\pm 1}\Psi_0$ with $J\neq J'$. 
This does not present any difficulties, however,  because the eigenvalues of Eq.~(\ref{tight binding s-wave}) are real.

\begin{figure*}[htbp]
\centering
\includegraphics[scale=0.54]{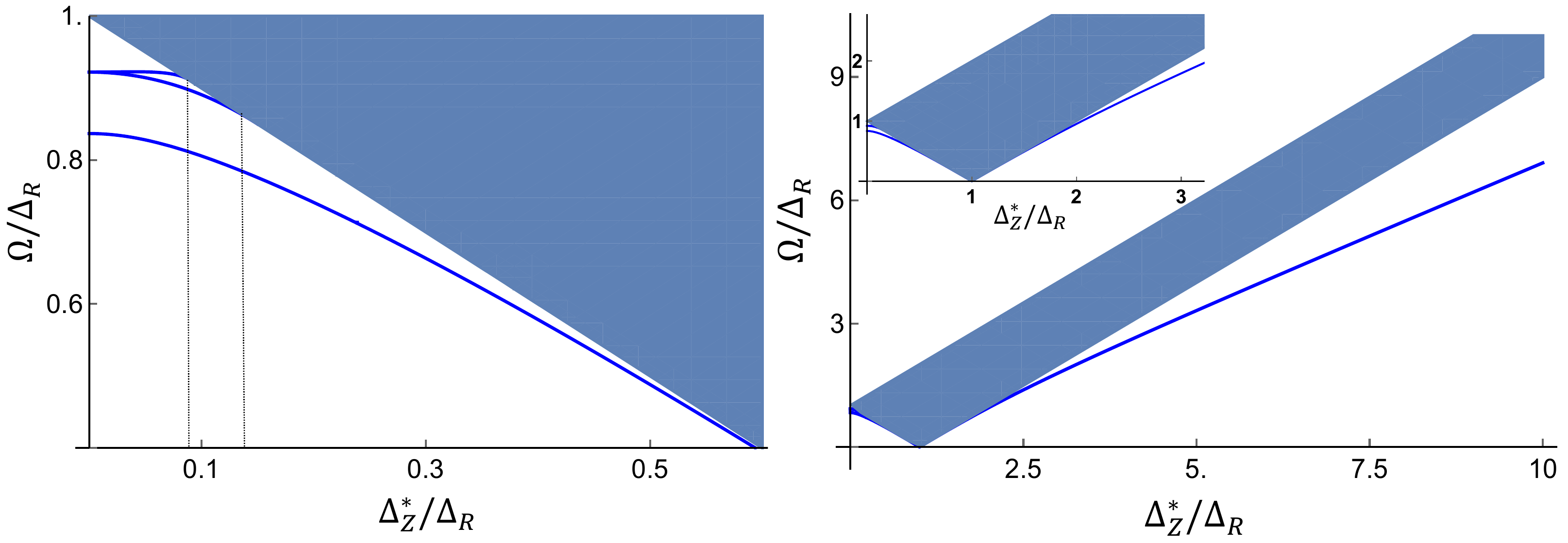}
\caption{\label{R+B_s_wave} Collective modes of a FL with Rashba SOC and in the presence of the in-plane  magnetic field. The Landau function is taken in the $s$-wave approximation: $F^a(\theta)=F_0^a=-0.3$. $\Delta_Z^*$ is the (renormalized) Zeeman energy and $\Delta_R$ is the Rashba energy splitting. Left: Weaker magnetic fields ($\Delta_Z^*<\Delta_R$). Right: Stronger magnetic fields ($\Delta_Z^*>\Delta_R$). Inset: Same as in main panels for a wider range of fields.}
\end{figure*}

Equation (\ref{tight binding s-wave}) can be solved by a slight modification of the standard method for finding the bound states on 1D lattices.\cite{kittel} Namely, we choose wavefunctions $\psi_0$ and $\psi_{\pm 1}$ as independent  variables. Starting from sites $m=\pm 2$, we assume that the wavefunction of the bound state decreases exponentially with $m$, i.e., 
\beq{\psi}_{\pm (|m|+2)} = e^{-(|m|+1)\lambda}
{\psi}_{\pm 1}
\label{Ansatz}
\eeq
 with $\text{Re}\lambda>0$. It is worth pointing out that eigenstates are localized in the angular-momentum space rather than real space. The localization radius of $\psi_{m}$ defines the harmonic content of a given collective mode, i.e., a state localized within $1/\text{Re}\lambda$ around $m=0$ contains effectively $m\lesssim 1/\text{Re}\lambda$ first harmonics.
 Applying Eq.~(\ref{Ansatz}) to any three nearest neighbors of the undistorted lattice, we obtain a relation between $\Omega^2$ and $\lambda$
\beq
\Omega^2=\Delta_{R}^{2} + \Delta_{Z}^{*2}-2\Delta_R\Delta_Z^*\cosh\lambda
\eeq
or
\begin{equation}
e^{-\lambda} = \frac{(1+x^2-y^2) \pm \sqrt{(1+x^2-y^2)^2 - 4x^2}}{2x},\label{lambda}
\end{equation}
where $y = \Omega/\Delta_R$ and $x = \Delta_Z^*/\Delta_R$. (To simplify the formulae, we will assume that $\Delta_R>0$.) 
Substituting $m=0,\pm 1$ into Eq.~(\ref{tight binding s-wave}) and using Ansatz (\ref{Ansatz}) to exclude $\psi_{\pm 2}$, we obtain a closed system for $\psi_0$ and $\psi_{\pm 1}$
\begin{widetext}
\begin{equation}
\label{s-wave matrix equation}
\left[\begin{array}{ccc}
1+x^2-y^2+\frac{1}{2} F_0^a -xe^{-\lambda} & -x\left[1+\frac{1}{2} F_0^a (3+F_0^a)\right] & 0 \\ 
-x(1+\frac{1}{2} F_0^a) & 1+x^2-y^2+ F_0^a \left[1+(2+F_0^a)x^2\right] & -x(1+\frac{1}{2}F_0^a) \\ 
0 & -x\left[1+\frac{1}{2} F_0^a (3+F_0^a)\right] & 1+x^2-y^2+\frac{1}{2} F_0^a -xe^{-\lambda}
\end{array} 
\right]
\left[\begin{array}{c}
{\psi}_{-1}  \\ 
{\psi}_0  \\ 
{\psi}_1
\end{array} 
\right]=0.
\end{equation}
\end{widetext}
Equating the determinant of this system to zero and using Eq.~(\ref{lambda}), we obtain a transcendental equation for eigenfrequencies. Although equation does have an analytic solution, its explicit form is quite lengthy,
and we delegate it to Appendix \ref{app:s_wave}, focusing here on Fig.~\ref{R+B_s_wave} obtained by plotting the results in Eq.~(\ref{collective_modes_swave}) of that Appendix.

The inset in Fig.~\ref{R+B_s_wave}b shows the spectrum for a broad range of the magnetic field, while panels a) and b) focus
on the regions of weak and strong magnetic fields, correspondingly. 
A prominent feature of the spectrum is the gap-closing point, $\Delta_Z^*=\Delta_R$, at which the continuum extends all the way down to zero energy leaving no room for collective modes. 
This happens when the spin-split Fermi surfaces touch and thus a spin-flip excitation costs no energy. To the left of this point, for $\Delta_Z^*<\Delta_R$, there are up to three collective modes which, at $\Delta_Z^*=0$ coincide with chiral spin modes, considered in Sec.~\ref{sec:RSOC}. The frequencies of these modes at $\Delta_Z^*=0$ are given by  Eqs.~(\ref{om0}) and (\ref{om1}) with $F_1^a=F^a_2=0$, i.e., 
\bse
\bea
\Omega_0&=&\Delta_R\sqrt{1+F_0^a}\label{om0s}\\
\Omega_{+1}&=&\Omega_{-1}=\Delta_R\sqrt{1+F_0^a/2}.\label{om1s}
\eea
\ese
 A finite magnetic field lifts the degeneracy of the $m=\pm 1$ modes which now disperse with $\Delta_Z^*$, as shown in  Fig.~\ref{R+B_s_wave}, left. At some critical values of the field, the $m=\pm 1$ modes run into the continuum, while the $m=0$ mode disperses all the way down to the gap-closing point. To the right of this point, for $\Delta_Z^*>\Delta_R$, there is only one mode which approaches asymptotically the Silin-Leggett mode in the limit $B\to\infty$. The frequency of this mode is given by the {\em bare} Zeeman energy, in agreement with the Kohn theorem.\cite{Kohn:1961,yafet:1963}
\subsection{Beyond $s$-wave approximation for the Landau function}
\label{sec:beyond_s}
In the previous section, we focused on the $s$-wave approximation for the Landau function. In this section, we analyze a more general case, beginning
with the Landau function which contains both the $m=0$ and $m=1$ harmonics ($s+p$--wave approximation):
\bea
F^a(\theta)=F_0^a+2F_1^a\cos\theta.\label{sp}
\eea

\begin{figure*}
\centering
\includegraphics[scale=0.54]{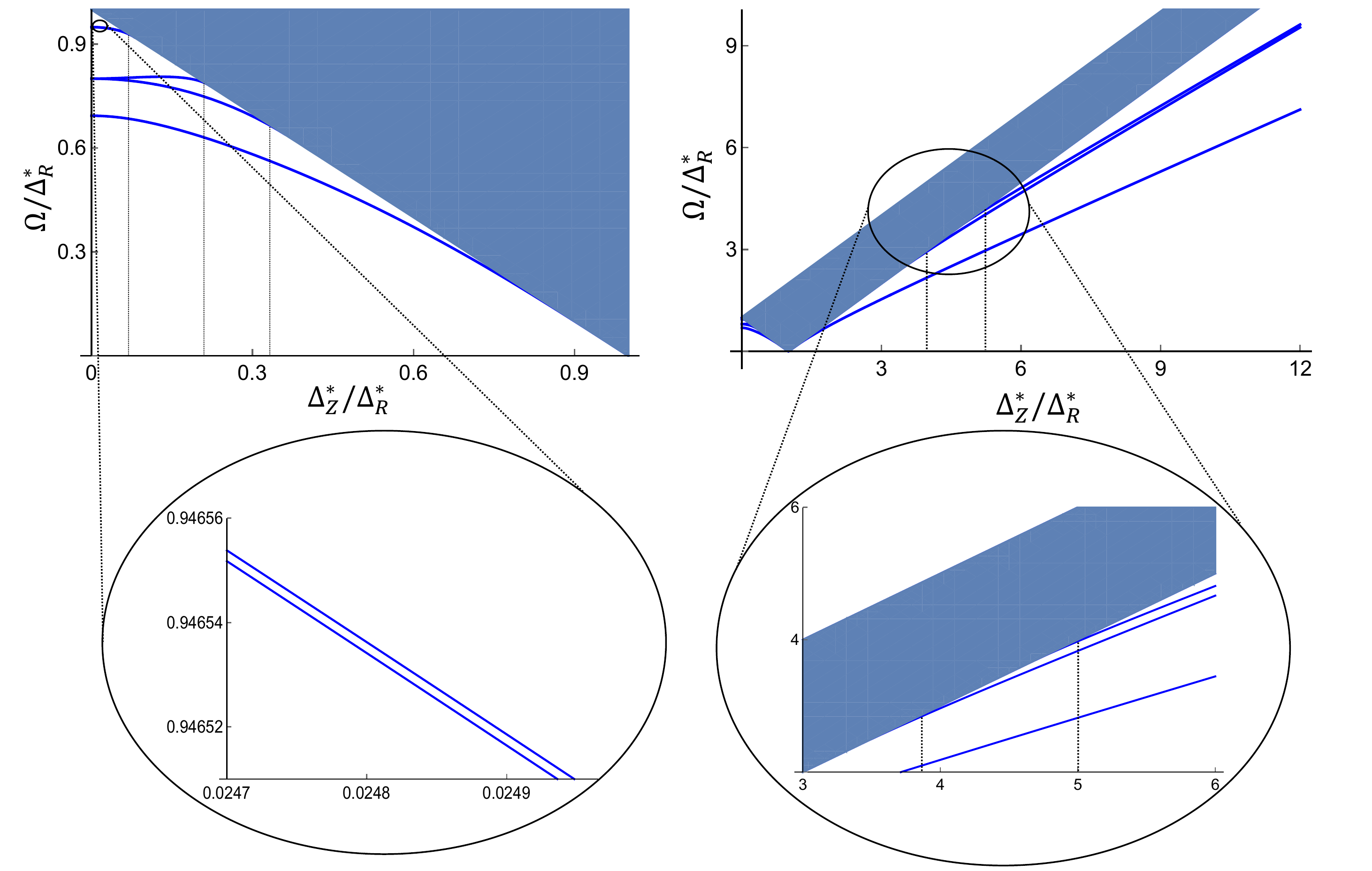}
\caption{\label{s_and_pwave} Collective modes of a FL with Rashba SOC and in the presence of the in-plane magnetic field. The Landau function is taken in the  $s+p$ approximation, Eq.~(\ref{sp}) . In the main panels,
$F_0^a = -0.4$ and $F_1^a = -0.2$. The zooms show two almost degenerate higher energy modes for $F_0^a = -0.4$ and $F_1^a = -0.4$. The magnitude of $F_1^a$ was increased to resolve the splitting.}
\end{figure*}

In  the effective lattice model, this case corresponds to five impurities
on sites $m=0,\pm 1,\pm 2$.  The spectrum of the collective mode for this case is plotted in Fig.~\ref{s_and_pwave}. There are five chiral spin resonances in zero magnetic field, which evolve into up to five collective modes as $\Delta_Z^*$ is increased up $\Delta_R^*$ (the left panel). A zoom below the main panel emphasizes a (numerically) small but finite splitting of the two highest-energy modes. As the field increases, four out of the five modes run into the continuum, following the same mechanism as for three-impurity case discussed in Sec.~\ref{sec:three}. On the opposite side of the spectrum, for $\Delta_Z^*\to\infty$, we now have two Silin-Leggett modes. At finite but small ratio of $\Delta_R^*/\Delta_Z^*$, the higher-energy Silin-Leggett mode splits into two, both of which run into the continuum at some critical values of $\Delta_R^*/\Delta_Z^*$. The remaining low-energy mode grazes the continuum and touches it at the gap-closing point. The same happens to the low-energy mode approaching the gap closing point from weak-field side. Details of the computational procedure can be found in Appendix \ref{app:s_p_wave}.

\begin{figure*}
\centering
\includegraphics[scale=0.55]
{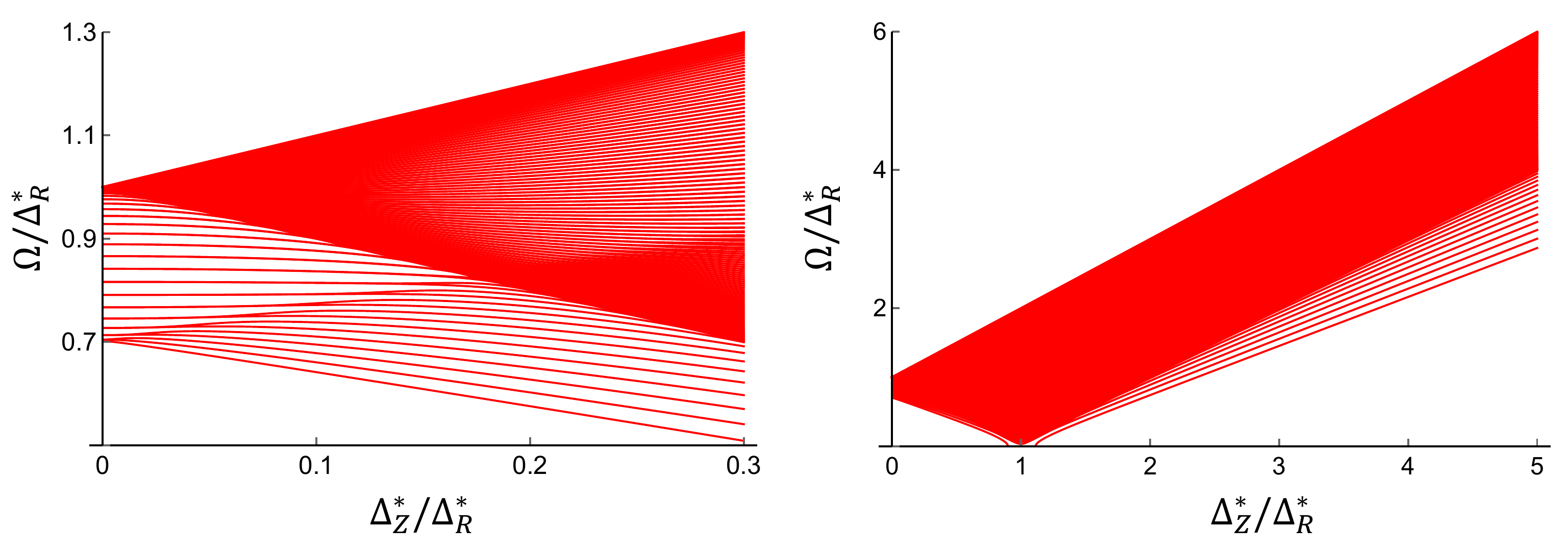}
\caption{\label{beyond_swave} Collective modes of a FL with Rashba SOC and in the presence of the in-plane magnetic field for a model form of the Landau function: $F_m^a = F_0^a \exp(-m^2/m_0^2)$ with $F_0^a=-0.3$ and $m_0 = 10$.}
\end{figure*}
If the Landau function contains all harmonics, one has to resort to numerical diagonalization of Eq.~(\ref{tight binding}) for a particular form of the Landau function. 
We choose an artificial but physically reasonable model with $F_m^a = F_0^a e^{-m^2/m_0^2}$. Numerical results for $F_0^a=-0.3$ and $m_0=10$ are shown in Fig.~\ref{beyond_swave}.  The spectrum is denser at lower energies because the harmonics of the Landau function with $m<m_0$ are close to $F_0^a$.  Equation (\ref{freqR}) shows that the frequencies of such modes at $\Delta_Z=0$ are close to the bare Rashba splitting, $\Delta_R$.  At the same time, the continuum at $\Delta_Z=0$ corresponds to a single energy equal to the renormalized Rashba splitting  $\Delta_R^*=\Delta_R(1+F_1^a)\approx \Delta_R/(1+F_0^a)>\Delta_R$. The modes with $m>m_0$ fill in the gap between $\Delta_R$ and $\Delta_R^*$. Although the collective mode spectrum is very dense, the modes remain discrete as long as $m_0$ is finite. In the effective lattice language, the system is equivalent to an alloy.  The central region of this alloy, $-m_0\lesssim m \lesssim m_0$, is occupied by impurities of comparable strength. Outside the central region, there are semi-infinite domains of weaker impurities whose strengths decrease rapidly away from the center.

\section{Physical interpretation \\ within the effective lattice model} 
\label{sec:interp}
\subsection{Effective tight-binding model}
Even in the simplest case of the $s$-wave approximation for the Landau function, the spectrum of the collective modes shown in 
Fig.~\ref{R+B_s_wave} is fairly complex and exhibits a number of distinct features. Namely, the two higher-energy modes in the region $\Delta_Z^*/\Delta_R<1$ merge with the continuum at certain values of the magnetic field; the lowest-energy mode runs into the continuum precisely at $\Delta_Z^*/\Delta_R=1$; there is only one mode for $\Delta_Z^*/\Delta_R>1$. The goal of this section
is to provide a transparent physical interpretation of these features  within the effective lattice model.

The effective tight-binding model corresponding to Eq.~(\ref{tight binding s-wave}) is depicted graphically in Fig.~\ref{chiral}. Equation (\ref{tight binding s-wave}) is an eigenvalue problem for the {\em square} of the frequency, and its RHS contains the squares of the various energy scales. To make an analogy with the tight-binding model complete, we will be referring to quantities with units of [energy]$^2$ simply to as ``energies''. In this way, $\Omega^2$ becomes the energy of the bound state $E$ while the  energy of hopping between normal sites is  
\beq J=\Delta_R\Delta_Z^*.\label{J}
\eeq
 The potential energies  on the defective sites will be measured relative to the on-site energy of undistorted lattice, $\Delta_R^2+\Delta_Z^{*2}$.   Then the potential energies on the defective sites $m=0$ and $m=\pm 1$ are given by
\bea
U_0&=&F_0^a\Delta_R^2+F_0^a(2+F_0^a)\Delta_Z^{*2}\;\text{and}\nn\\
U_1&=&(F_0^a/2)\Delta_R^2,\label{U01}
\eea
correspondingly.
Finally, the hopping amplitudes between the defective sites ($m=0$ and $m=\pm 1$) are
\bea
t&=&\left(1+\frac{F_0^a}{2}\right)J,\;\text{for}\; 0\to \pm 1;\nn\\
t'&=&\left[1+\frac{F_0^a}{2}(3+F_0^a)\right]J,\;\text{for}\; \pm 1\to 0.\label{ttp}
\eea
For $F_0^a<0$, the impurities are {\em attractive}, i.e., $U_0,U_1<0$, while the defective bonds are {\em weaker} than the normal ones, i.e., $t,t'<J$.

\subsection{Simplified lattice models}
To explain every detail of the spectrum in Fig.~\ref{R+B_s_wave}, one needs to take into account all of the elements listed above. However,  certain  features can be understood by considering simplified versions of the tight-binding model, which is what we are going to do in the next sections as well as in Appendix \ref{TB}.
\subsubsection{Three attractive impurities}
\label{sec:three}
\begin{figure}[htbp]
\centering
\includegraphics[scale=0.35]{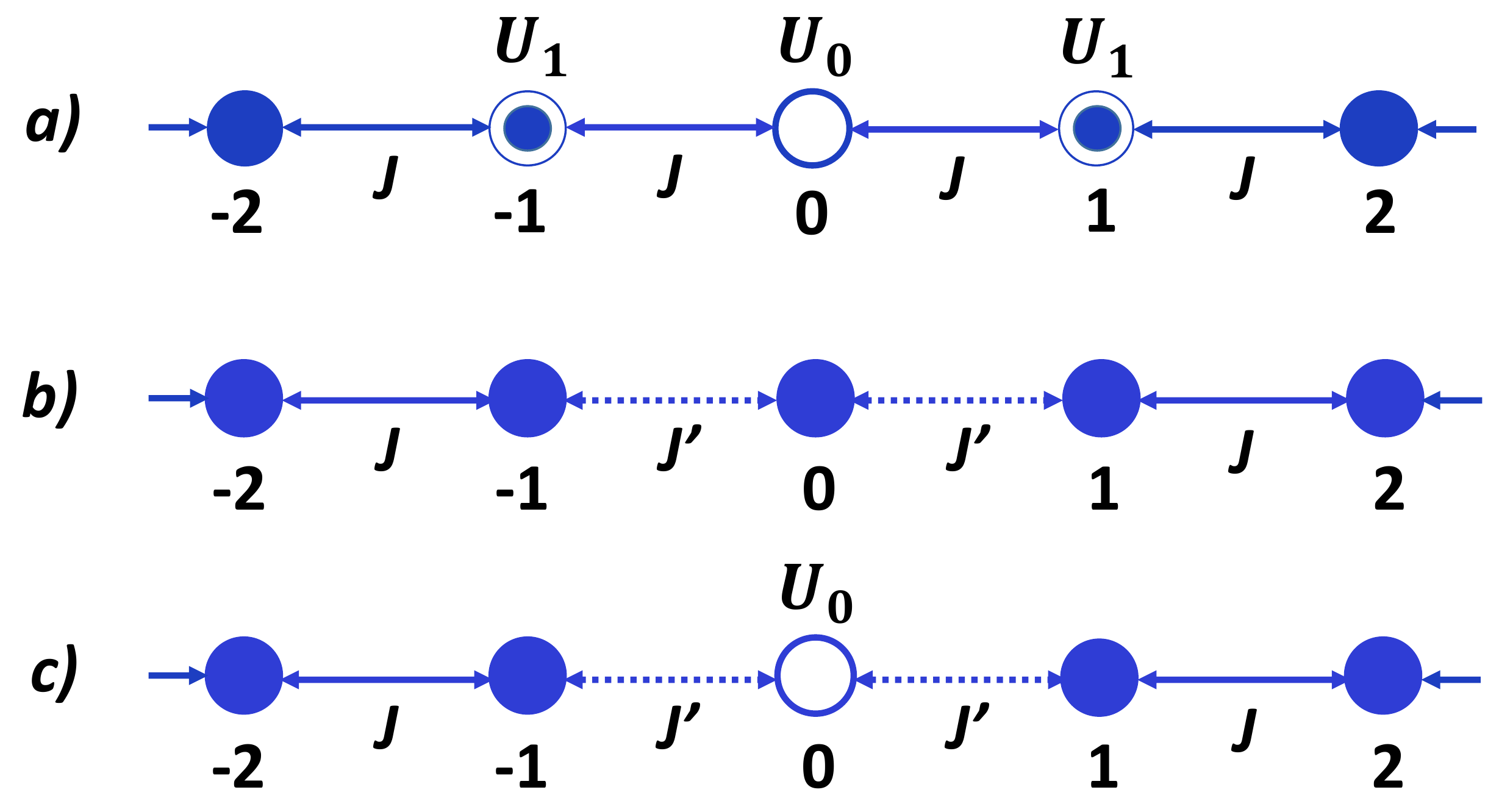}
\caption{\label{three_on-site}One-dimensional tight-binding models.
a)  Three attractive impurities ($U_0,U_1<0$) at $m=0$ and $m=\pm 1$. b) Two defective bonds between sites $m=0$ and $m=\pm 1$ with hopping amplitudes $J'$.  c) A single attractive impurity at $m=0$ and two defective bonds between sites $m=0$ and $m=\pm 1$ with hopping amplitudes $J'$. }
\end{figure}

The merging of the two higher-energy modes with the continuum--can be understood qualitatively by ignoring bond defects, i.e., by setting $t=t'=J$. In this case, we have a tight-binding model with normal bonds between all sites and with three impurities on sites $m=0,\pm1$, see Fig.~\ref{three_on-site}a. It is also convenient to measure all energies in units of $J$, i.e., to set $J=1$.

For an even simpler case of a single on-site defect, it is well-known that there always exists a bound state located either below (for $U_0<0$)  or above (for $U_0>0$) the conduction band.
In the context of semiconductors, these states are known as donors and acceptors, i.e., the bound states of electrons and holes, correspondingly. In the tight-binding model, the states with energies below the inflection point 
are electron-like (with positive effective mass) while the states with energies above the inflection point are hole-like (with negative effective mass), therefore, the donor and acceptor states occur in this case as well. Since our impurities are attractive, we will focus on this case from now on.

\begin{figure}[htbp]
\centering
\includegraphics[scale=0.4]{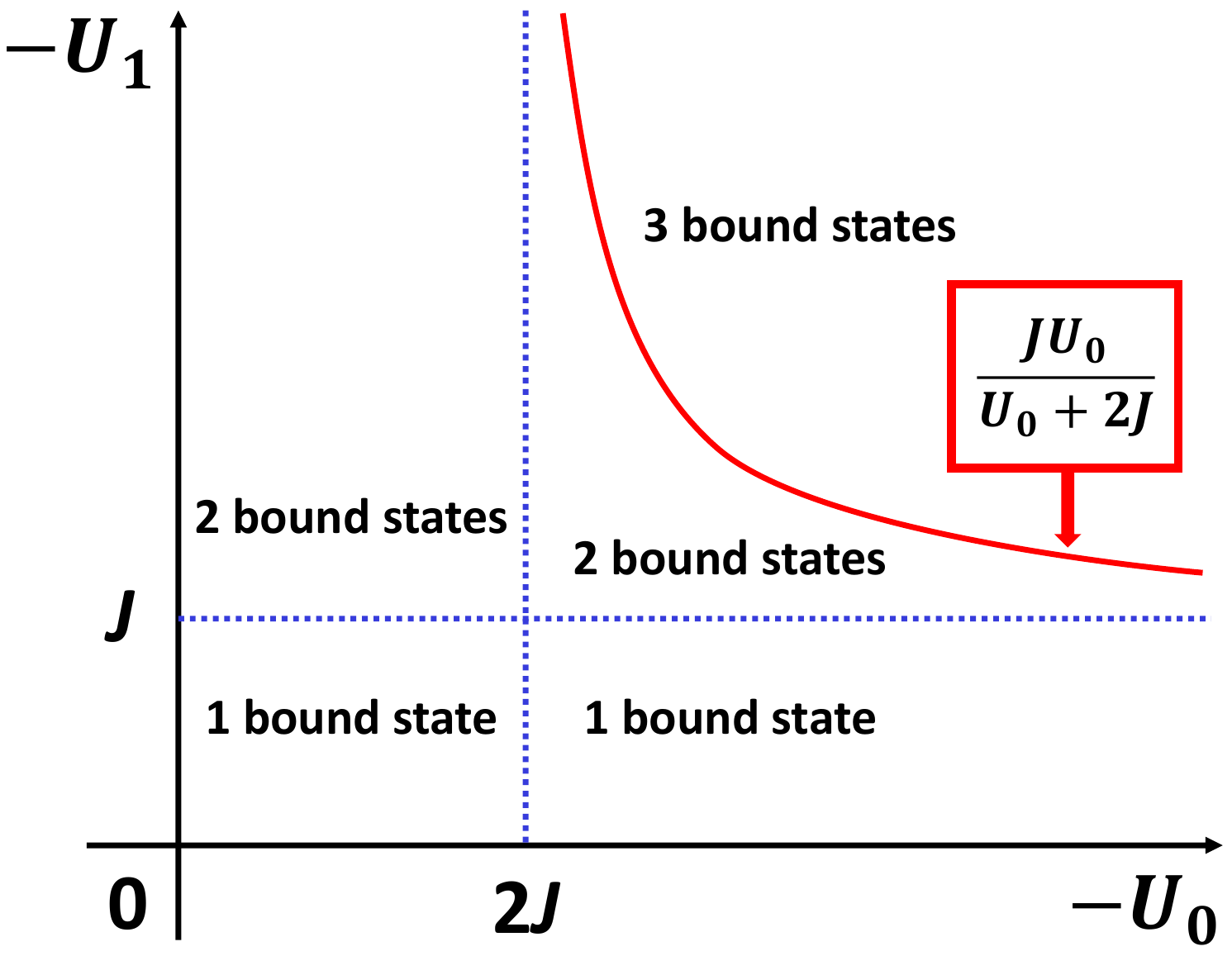}
\caption{\label{phase-diagram}
Phase diagram for bound states in a one-dimensional tight-binding model with three attractive on-site impurities, as shown in Fig.~\ref{three_on-site}a.}
\end{figure}

Given that 
a single impurity has one bound state, it is natural to expect that three impurities will have up to three bound states. In the continuum limit, the three-impurity complex corresponds to a 1D potential well of {\em finite} width ($a$) and depth ($U$), which has at least one bound state  but may also have two, three, etc. states,  if the product $-Ua$ exceeds some critical values. 
The lattice case is analyzed in Appendix \ref{sec:three_defects} and summarized in the phase diagram shown in Fig.~\ref{phase-diagram}. There is indeed at least one and up to three bound states, depending on the impurity strength. The lowest energy eigenstate is of even parity ($\psi_1=\psi_{-1}$), the next one is odd ($\psi_1=\psi_{-1}$), and the highest one is again even.

Our original problem corresponds to a tight-binding model with parameters given by  Eqs.~(\ref{J}), (\ref{U01}), and (\ref{ttp}).
In the limit $\Delta_Z^*\ll \Delta_R$, the potential energies of all the three impurity sites are of the order of $\Delta_R^2$, which is much larger than the bandwidth $2J=2\Delta_R\Delta_Z^*$. Thus we have three strong impurities with the maximum number of bounds states, which is equal to three. As $\Delta_Z^*$ is increased, the bandwidth increases as $\Delta_Z^*$ but the potential energies increase only as $\Delta_Z^{*2}$. Therefore, the impurities get relatively weaker (compared to the bandwidth), and we lose first the highest and then next-to-highest-energy bound state, when the ratio of the potential energy to the bandwidth falls below some critical values. This explains why two out of the three collective modes merge with the continuum at certain values of $\Delta_Z^*/\Delta_R<1$. 

\subsubsection{Competition between on-site and bond defects} 
\label{sec:comp}
The three-impurity model is unable to explain an interesting feature of the spectrum in Fig.~\ref{R+B_s_wave}: the lowest energy collective mode approaches the continuum either from the left (panel {\em a} ) or from the right (panel {\em b}) and touches the continuum at one particular point, where $\Delta_Z^*=\Delta_R$. From the lattice point-of-view, it means that  a 1D tight-binding model does not have a bound state for a certain choice of parameters, which cannot happen if only on-site defects are present. The reason for such a behavior is the competition between impurities and adjacent defective bonds, which we discuss below. 

To understand how does this competition work, we first consider a toy model with no impurities but 
with two defective bonds (Fig.~\ref{three_on-site}b).
As shown in Appendix \ref{sec:bondsA},  a bound state occurs  in this case only if the defective bonds are {\em stronger} than the normal ones, i.e., $J'>J$. (There are actually two bound states: one above and one below the conduction band.) 

\begin{figure}[htbp]
\centering
\includegraphics[scale=0.30]{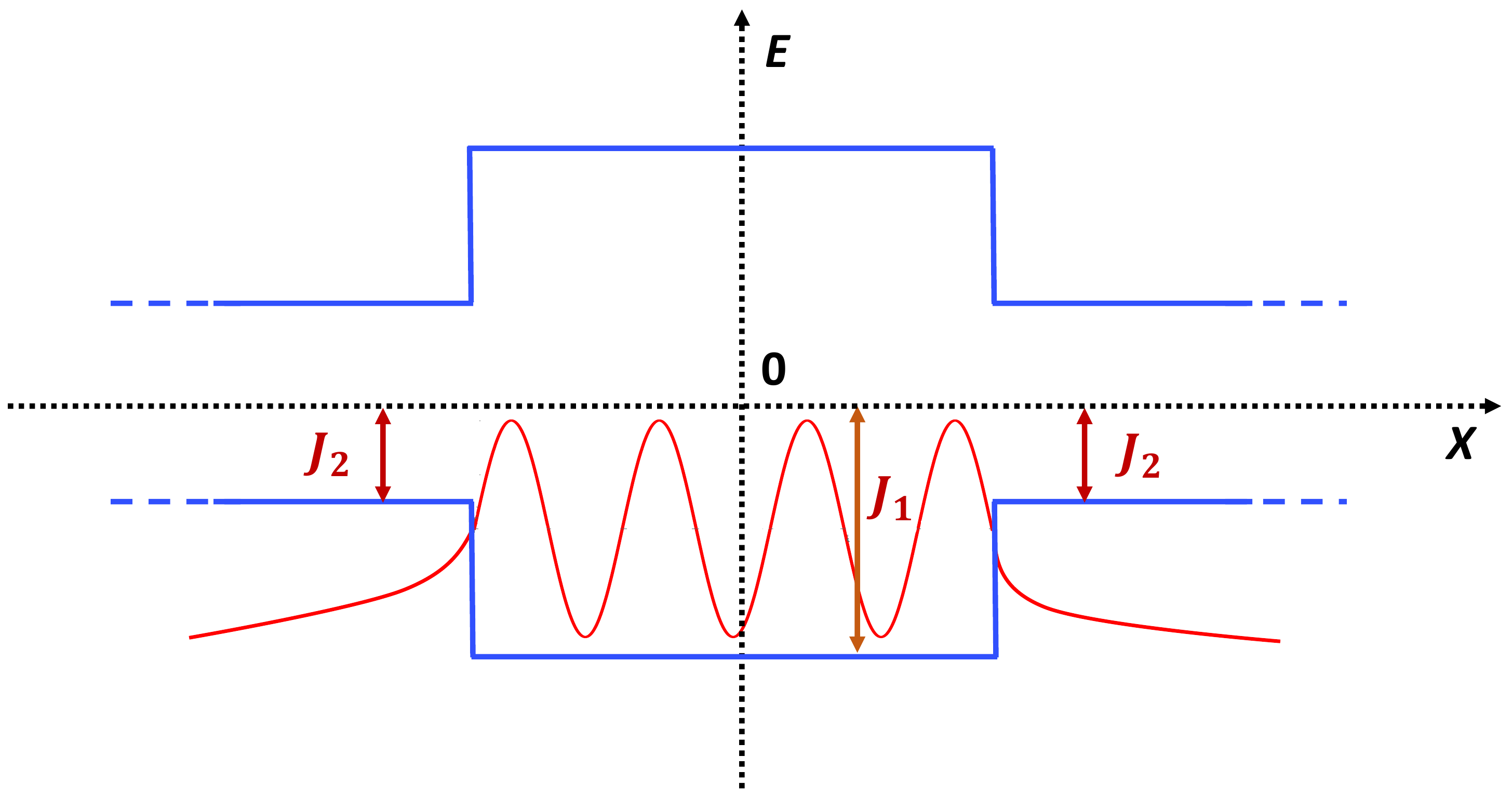}
\caption{\label{Band} A bound state in a junction between a wide-band and narrow-band materials. Electrons with energies $J_2<|E|<J_1$ are confined to the central region.}
\end{figure}

The difference between the cases of weaker and stronger defective bonds can be understood by going to the continuum limit, where  a local change in $J$ leads to spatial variations in both the bandwidth and effective mass.  The latter does not give rise to a bound state by itself. Indeed, the Schroedinger equation  with a step-like variation in the mass does not have an evanescent solution, which means that there are no bound states. On the contrary, a local variation in the bandwidth  gives rise to a bound state, only if the band is wider in the central region. 
Indeed, neglecting the spatial variation of the effective mass, the Schroedinger equation corresponding to the tight-binding model reads 
\begin{equation}
\label{continuum}
\left[E+2J(x)\right] \psi(x) = - \frac{1}{2m}\frac{d^2 \psi(x)}{dx^2},
\end{equation}
where $m=1/2J$ (the lattice constant is set to unity). An electron with energy $E>-2J(x)$ is free to move, while an electron with energy  $E<-2J(x)$ is localized.
Now consider an interface between two materials with hopping amplitudes  $J_1$ and $J_2$, to the left and to the right of the interface, correspondingly, see Fig.~\ref{Band}. An electron with energy in the interval $-2J_1<E<-2J_2$ is free to move in the left half-space but cannot propagate into the right half-space.  The condition $-2J_1<E<-2J_2$ implies that $J_1>J_2$, i.e., that the band is wider in the central region. Adding another interface leads to the formation of a bound state within the material with larger $J$, i.e., with a wider band. A similar reasoning works for the states with energies near the top of the band (holes). 
This effect is dual to a bound state in a narrow-gap semiconductor sandwiched between two wide-gap semiconductors.\cite{vasko}

Next, we add a single attractive impurity to the model and connect it by two {\em weak} bonds to the rest of the lattice, as shown in 
Fig.~\ref{three_on-site}c. Because no bound state exists in the presence of weak bonds only, there should be a competition between the impurity, which would like to have a bound state, and weak bonds, which do not. As the bonds get weaker, the bound state becomes shallower until, at some critical value of $J'$, it merges with the band. (As shown in Appendix \ref{sec:1i2b}, this happens if $|U_0|\leq 2J$ and at the critical bond strength of $|J_c|=J\sqrt{1-|U_0|/2J}$.)
\subsection{Weak magnetic fields: $\Delta_Z^*<\Delta_R$}
Since the defective bonds in our case are indeed weak [see Eq.~(\ref{ttp})], we now understand qualitatively the mechanism by which the lowest-energy collective mode runs into the continuum. However, it does not explain why this happens precisely
at $\Delta_Z^*=\Delta_R$ rather than at some arbitrary value of  $\Delta_Z^*$. To understand this, we need to return to the full tight-binding model in Fig.~\ref{chiral}a, which contains all elements
of the original problem, i.e., three impurities, and two defective and chiral bonds with $t\neq t'$, with parameters exactly as in Eqs.~(\ref{J}), (\ref{U01}), and (\ref{ttp}).

In the case, the eigenvalue problem is reduced to a $3\times 3$ system of equations
\begin{equation}
\label{TB_chiral}
\begin{split}
E \psi_0 &= U_0 \psi_0 - t (\psi_{-1} + \psi_{1}), \\
E \psi_{\pm 1} &= U_1 \psi_{\pm 1} - t' \psi_0 - J\psi_{\pm 1}e^{-\lambda}.
\end{split}
\end{equation}
where we used {\em Ansatz} (\ref{Ansatz}) to eliminate $\psi_{\pm 2}$ in favor of $\psi_{\pm 1}$. The same Ansatz, being applied to any three adjacent sites of the undistorted lattice yields $E=-2J\cosh\lambda$.

 Selecting even- and odd-parity solutions, we obtain equations for the corresponding eigenvalues
\bes
\bea
\label{toy}
&&\left( E - U_1 + Je^{-\lambda}\right) \left(E - U_0\right) = 2 t t',\;\text{even};\label{up}\\
&&E -U_1 + Je^{-\lambda}=0,\;\text{odd}.\label{down}
\eea
\ees
The lowest energy bound state must be of even parity. Therefore, the condition for its disappearance must follow from Eq.~(\ref{up}). The bound state coincides with the band edge if $E=-2J=-2\Delta_R\Delta_Z^*$. Substituting this value of $E$ along the rest of the parameters from Eqs.~(\ref{U01}) and (\ref{ttp}) into Eq.~(\ref{up}), we find that one of its two solutions is $\Delta_Z^*=\Delta_R$, which is indeed the gap-closing point. The second solution is $\Delta_Z^*=-F_0^a/[2(2+F_0^a)]\Delta_R$. This is precisely the point where the highest-energy collective mode, which is also of even parity, runs into the continuum (cf. Fig.~\ref{R+B_s_wave}a). Finally, substituting 
the same parameters into Eq.~(\ref{down}), we find that the odd-parity collective mode runs into the continuum at $\Delta_Z^*=-F_0^a\Delta_R/2$. These analytic results  are in precise agreement with the exact solution in Appendix \ref{app:s_wave}.

It is worth pointing out that the disappearance of the collective mode precisely at the point where $\Delta_Z^*=\Delta_R$ is not coincidental. The parameters of the tight-binding model are derived from the FL kinetic equation
and thus bear the information about the symmetries of the underlying model, i.e., the $SU(2)$ symmetry of the original FL (in the absence of  SOC) and the $C_{\infty v}$ symmetry of the Rashba Hamiltonian. It is thus no accident that the bound state of the effective lattice model disappears precisely at the point where the gap in the continuum closes.

\subsection{Strong magnetic field: $\Delta_Z^*>\Delta_R$.}
As Fig.~\ref{R+B_s_wave} shows, there is only one bound state  for $\Delta_Z^*>\Delta_R$. It is easier to understand this case starting from the limit of $\Delta_Z^*\to\infty$, where the Rashba term is negligibly small. In this limit,
hopping disappears and we have decoupled sites with energies $\Delta_Z^{*2}$ on all sites but at $m=0$, where the on-site energy is given by bare Zeeman splitting, $\Delta_Z^2$. This energy gives the frequency of the Silin-Leggett collective mode, while sites with $m\neq 0$ form a continuum at $\Delta_Z^{*}$.  It is intuitively obvious that small $\Delta_R$ cannot change the picture qualitatively: we must still have just one bound state with a renormalized frequency. Indeed, expanding the exact solution in Eq.~(\ref{collective_modes_swave}) for $\Delta_Z^* \gg |\Delta_R|/|F_0^a|$, we obtain
\begin{equation}
\Omega_L=\Delta_Z + \frac{(2+3F_0^a)(1+F_0^a)}{4F_0^a} \frac{\Delta_R^2}{\Delta_Z} +\dots,
\label{largeZ}
\end{equation}
which coincides with the RPA result of Ref.~\onlinecite{maiti2016} upon replacing the dimensionless coupling constant $u$ by $-F_0^a$. Notice that the Kohn theorem, which states that the Larmor frequency is not affected by the electron-electron interaction, holds for the leading term in Eq.~(\ref{largeZ}) but is violated already for the first correction due to the presence of SOC.

As $\Delta_Z^*$ decreases, the impurity at $m=0$ remains relatively strong (the ratio of its potential energy to the bandwidth is on the order of $\Delta_Z/\Delta_R\gg 1$), while the impurities at $m\pm 1$ remain weak (the corresponding ratio is on the order of $\Delta_R/\Delta_Z\ll1$). In the case, there is only one bound state (cf. phase diagram in Fig.~\ref{phase-diagram}). As $\Delta_Z^*$ becomes comparable to $\Delta_R^*$, the strengths of all the three impurities become comparable both to each other and to the bandwidth, and more bound states may appear. However, as we explained above, our model is fine-tuned by the choice of parameters corresponding to the original FL kinetic equation, and with this choice there is only one bound state for $\Delta_Z^*>\Delta_R$, which touches the continuum at $\Delta_Z^*=\Delta_R$.

To conclude this section, we re-iterate that all the features of the collective-mode spectrum are accounted for in the effective lattice description.

\section{Fermi liquid with Rashba and Dresselhaus spin-orbit coupling}
\label{sec:R+D}

In this section, we consider the case when both Rashba and Dresselhaus  SOCs are present but there is no external magnetic field. 
 This situation is similar to that of the magnetic field and Rashba SOC, considered in Sec.~\ref{sec:Rashba+B}, except for hopping is now of the next-to-nearest-neighbor type. This will lead to important differences in the collective-mode spectrum. 
 
 Setting $\Delta_Z^*=0$ in Eqs.~(\ref{u1_m} -\ref{u3_m}), reducing the system of equations  to a single equation for $u_1^m$  as before, and introducing the Bloch wavefunction via Eq.~(\ref{vm}), we obtain the effective tight-binding model as \begin{widetext}
\bea
\label{tight binding dresselhaus}
\Omega^2\psi_m &=& \big(\Delta_{R}^{*2}+\Delta_{D}^{*2} \big)  \big(1+F_m^a \big) \bigg[ 1+\frac{1}{2} \big( F_{m+1}^a + F_{m-1}^a \big) \bigg]  \psi_m - \Delta_R^* \Delta_D^* \Big(1+F_{m-2}^a \Big) \Big(1+F_{m-1}^a \Big) \psi_{m-2} \nn\\
&&-\Delta_R^* \Delta_D^* \Big(1+F_{m+2}^a \Big) \Big(1+F_{m+1}^a \Big) \psi_{m+2}.
\eea
\end{widetext} 
The effective lattice is now bipartite: every even (odd) site is coupled to the nearest even (odd) sites  but there is no coupling between sites of different parity. It is convenient then to consider the lattice as being composed of two decoupled sublattices which contain only even or only odd sites, with nearest-neighbor hopping within each sublattice.
Introducing the sublattice wavefunctions as $\chi_l=\psi_{2l}$ and $\xi_l=\psi_{2l+1}$, we obtain two independent equations 
 \begin{widetext}
\begin{equation}
\label{sub}
\begin{split}
\Omega^2\chi_l = \big(\Delta_{R}^{*2}+\Delta_{D}^{*2} \big)&  \big(1+F_{2l}^a \big) \bigg[ 1+\frac{1}{2} \big( F_{2l+1}^a + F_{2l-1}^a \big) \bigg]  \chi_l - \Delta_R^* \Delta_D^* \Big(1+F_{2l-2}^a \Big) \Big(1+F_{2l-1}^a \Big) \chi_{l-1}  \\
&-\Delta_R^* \Delta_D^* \Big(1+F_{2l+2}^a \Big) \Big(1+F_{2l+1}^a \Big) \chi_{l+1},\\
\Omega^2\xi_l = \big(\Delta_{R}^{*2}+\Delta_{D}^{*2} \big)&  \big(1+F_{2l+1}^a \big) \bigg[ 1+\frac{1}{2} \big( F_{2l+2}^a + F_{2l}^a \big) \bigg]  \xi_l - \Delta_R^* \Delta_D^* \Big(1+F_{2l-1}^a \Big) \Big(1+F_{2l}^a \Big) \xi_{l-1}  \\
&-\Delta_R^* \Delta_D^* \Big(1+F_{2l+3}^a \Big) \Big(1+F_{2l+2}^a \Big) \xi_{l+1}.
\end{split}
\end{equation}
\end{widetext} 
The on-site energies of the ideal lattice are given by $\Delta_R^{*2}+\Delta_D^{*2}$, while the hopping amplitude is $J=\Delta_R^*\Delta^*_D$.

\begin{figure}[htbp]
\includegraphics[scale=0.31]{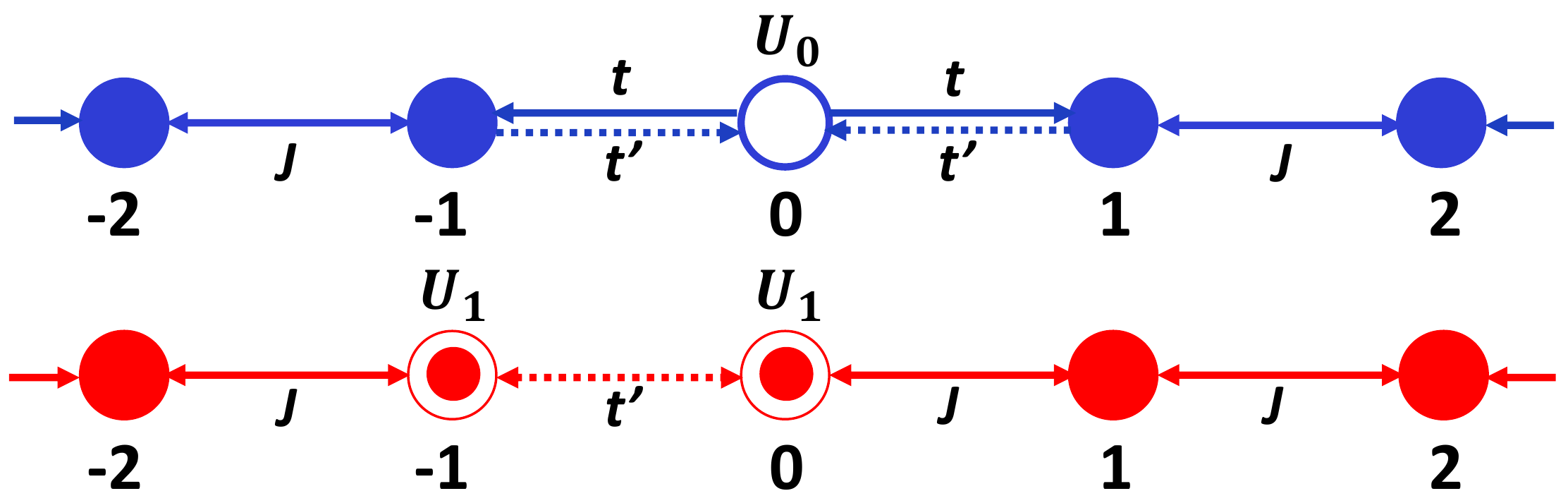}
\caption{\label{R+D_lattice} Effective lattice model for a FL with both Rashba and Dresselhaus spin-orbit couplings in the $s$-wave approximation.}
\end{figure}

The effective lattice model in the $s$-wave approximation for the Landau function is depicted graphically in Fig.~\ref{R+D_lattice}. The even sublattice (top) has a single attractive impurity at $l=0$. The bonds between the $l=0$ and $l=\pm 1$ sites are chiral ($t\neq t'$): the bond in the forward direction (from $0$ to $\pm 1$) is undistorted ($t=J$), while the bond in the backward direction (from $\pm 1$ to $0$) is weak ($t'<J$). If not for the bond defects, the even sublattice would have had a single bound state with an even-parity wavefunction. The odd sublattice has two impurities (on sites $l=-1$ and $l=0$) connected by a weak non-chiral bond. The odd sublattice can have up to two bound states, 
with even- and odd-parity wavefunctions. As in the case of Rashba SOC plus magnetic field (R+B), the maximum number of bound states is three. The difference  is  that, in the R+B case, two out of the three states are the ``extra" states, which occur only if impurities are sufficiently strong. As the magnetic field increases, the impurities get weaker (relative to the bandwidth), and these extra states merge with the continuum one by one.
In the Rashba plus Dresselhaus case, the three bound states come as a singlet from the even sublattice and a doublet from the odd sublattice. Only one of the components of the doublet is an extra state, which can merge with  the continuum. The remaining component of the doublet and the singlet are the lowest-energy bound states, which can only be eliminated by a competition between on-site and bond defects. Therefore, one should expect both these states to graze the continuum and touch it a special point, where the Rashba and Dresselhaus couplings compensate each other. 

\begin{figure*}[htbp]
\centering
\includegraphics[scale=0.54]{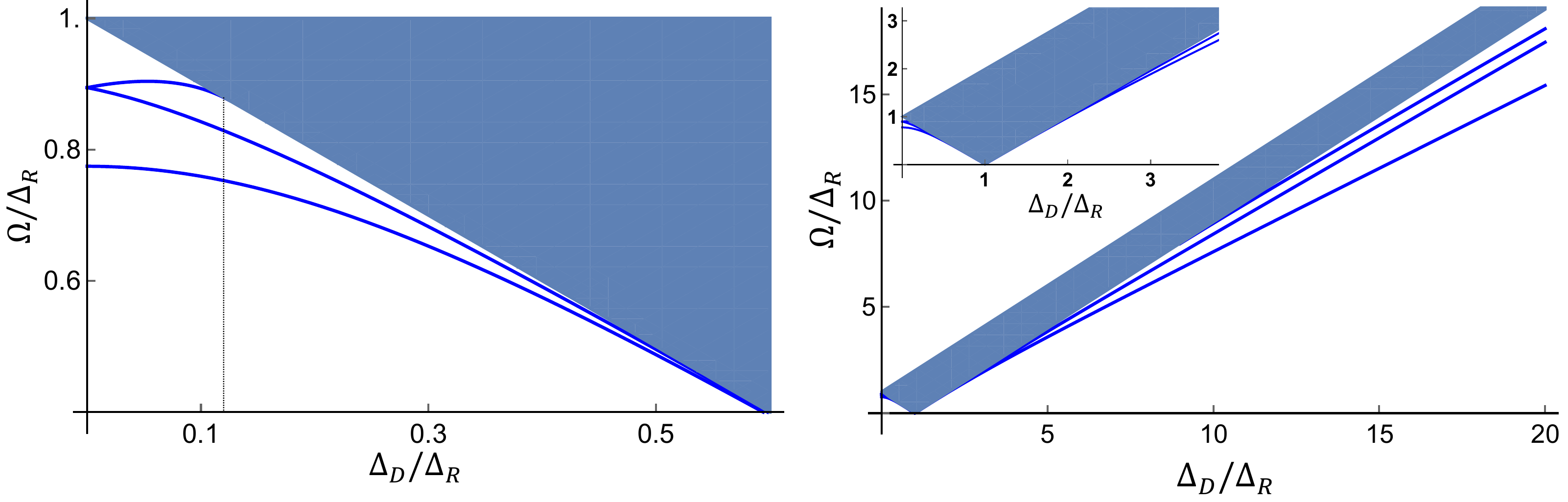}
\caption{\label{R+D_s_wave} Collective modes of a FL with Rashba and Dresselhaus spin-orbit couplings. The Landau function is taken in the $s$-wave approximation with $F_0^a = -0.4$. Left: Weaker Dresselhaus SOC ($\Delta_D<\Delta_R$). Right: Stronger Dresselhaus SOC ($\Delta_D>\Delta_R$). Inset: Same as in main panels for a wider range of Dresselhaus coupling.}
\end{figure*}

This qualitative picture is indeed confirmed by an exact solution of Eq.~(\ref{sub}), shown in Fig.~\ref{R+D_s_wave}. The spectrum is symmetric about the $SU(2)$-symmetric point, where $\Delta_R=\Delta_D$ (see Fig.~\ref{R+D_s_wave}, inset).\cite{schliemann:2003,bernevig:2006} There are up to three collective modes to each side of this point. The highest-energy mode runs into the continuum at some value of  $\Delta_R/\Delta_D$, while the two lower-energy modes graze the continuum and touch it at $\Delta_R=\Delta_D$. The conditions for touching can be derived in the same way as it was done in the previous section for the R+B case.

\section{Conclusions}
\label{sec:conc}
In conclusion, we showed that the quantum kinetic equation for a Fermi liquid  can be mapped onto an effective one-dimensional tight-binding model, in which the lattice sites correspond to the angular-momentum channels of the non-equilibrium part of the occupation number. In this mapping, the Rashba term plays the role of the on-site energy, while the Zeeman and Dresselhaus terms are responsible for nearest and next-to-nearest neighbor hopping between the sites of an ideal lattice, correspondingly. Consequently,  the continuum of spin-flip particle-hole excitations becomes the conduction band of the lattice model. The Fermi-liquid interaction, characterized by the harmonics of the Landau function, produces defects of both on-site and bond type. The collective modes correspond to the bound states produced by these defects. We showed that all the features of the collective-mode spectrum can be explained naturally within the lattice model as a result of the competition between on-site and bond defects.

Although we focused on a particular example of a partially spin-polarized Fermi liquid with spin-orbit coupling and considered only the spatially uniform ($q=0$) limit, we believe that mapping
onto an effective lattice model can be useful in other cases as well. For example, consider a kinetic equation describing the zero-sound modes in a neutral
two-dimensional Fermi liquid \cite{statphys}

\begin{equation}
(\omega - v_F^*q \cos\theta)u(\theta) = v_F^*q\cos\theta\int \frac{d\theta'}{2\pi} F^s(\theta-\theta')u(\theta'),
\end{equation}
where $v_F^*$ is the renormalized Fermi velocity. In the harmonic representation, the equation above reads
\bea
\omega u_m&=&\frac 12 v^*_Fq (u_{m+1}+u_{m-1})\nn
\\&&+\frac 12 v^*_Fq \left(F^s_{m+1}u_{m+1}+F^s_{m-1} u_{m-1}\right).
\eea
As before, we can view this equation as an effective tight-binding model. The $v_F^*q$ term induces hopping between nearest neighbors. In a non-interacting system, this hopping forms a band of width $v_Fq$ which corresponds to the continuum of particle-hole excitations with energies $0\leq\omega\leq v_Fq$ (we consider excitation with positive energies). The Fermi-liquid interaction plays the role of {\em bond} defects (there are no on-site defects in this case). For example, in the $s$-wave approximation, the bonds between $m=0$ and $m=\pm 1$ sites and defective and chiral: the hopping amplitude from $m=0$ to $m=\pm 1$ is the same as for an ideal lattice but that in the reverse direction is multiplied by a factor of $1+F_0^s$. As we showed in Sec.~\ref{sec:comp} and Appendix \ref{sec:bondsA}, a bond defect forms a bound state only if it is strong, i.e., its hopping amplitude is larger than that for an ideal lattice. Therefore, there should a bound state above the continuum if $F_0^s>0$. This bound state is nothing but the zero-sound wave with velocity
$v_F^*(1+F^s_0)/\sqrt{1+2F^s_0}$ (Refs.~\onlinecite{suhas:2005b,chubukov:2005b}).

It is also easy to understand within the lattice picture why a plasmon in a charged Fermi liquid never merges with the continuum but always stays above its boundary.\cite{physkin} This is so because a one-dimensional lattice with single type of defects (of the bond type in this case) must have at least one bound state which corresponds to the plasmon mode.

We believe that the lattice interpretation will be also useful for the analysis  of more complicated cases, e.g., of Raman spectroscopy of semiconductor heterostructures,\cite{perez:2013,perez:2015,perez:2016,karimi:2016,maiti:2017} which measures spatial dispersion of chiral spin waves formed in the presence of both Rashba and Dresselhaus types of spin-orbit coupling, and also of the in-plane magnetic field. 
\acknowledgments
We thank C. Batista, M. Imran, S. Maiti,  C. Reeg, and A. Rustagi for stimulating discussions. A part of this work was carried out at the Center for Non-Linear Studies of Los Alamos National Laboratory, which D. L. M. visited as an Ulam Scholar in 2015-2016.

\appendix
\section{One-dimensional tight-binding model with on-site and bond defects}
\label{TB}

In this Appendix, we present solutions of a number of one-dimensional tight-binding models with impurities and bond defects,
discussed in the main text.

\subsection{On-site defects only}
\subsubsection{Single on-site defect}
To set the stage, we review briefly the solution of the simplest model with a single on-site defect (``impurity''). Given that an impurity with potential energy $U_0$ occupies the $m=0$ site, the system of Schroedinger equations read
\bse
\bea
E \psi_0 &=& U_0 \psi_0 - \psi_{-1} - \psi_{1}, \label{A1a}\\
E \psi_{\pm(|m|+1)} &=& - \psi_{\pm |m|} - \psi_{\pm(|m|+2)} \label{A1b}.
\eea
\ese
(The hopping amplitude $J$ is set to unity.)
We assume that the bound-state wavefunction decays exponentially away from the central site: 
\bea
\psi_{\pm(|m|+1)}=\psi_{\pm |m|}\exp(-|m|\lambda)
\label{ansatz}
\eea
with $\text{Re}\lambda>0$.
Substituting this {\em Ansatz} into Eq.~(\ref{A1b}), we obtain \beq
E=-2\cosh\lambda\label{reg}
\eeq 
or
\begin{equation}
e^{-\lambda_{\pm}} = -\frac{E}{2} \pm \sqrt{\frac{E^2}{4}-1}.\label{el}
\end{equation}
 The root with the minus sign corresponds to  a bound state below the conduction band, because $\text{Re}\lambda_->0$ for $E<-2$.  Conversely, the root with the plus sign corresponds to a bound state above the conduction band, because $\text{Re}\lambda_+>0$ for $E>2$. Substituting these roots into Eq.~(\ref{A1a}), we obtain for the bound state energy
 \beq
 E=\text{sgn} U_0\sqrt{U_0^2+4}.
 \eeq
 For $E<-2$, the exponent $\lambda_-$ is real. Correspondingly, the wave function decreases with $m$ in a purely exponential manner: $\psi_m\propto \exp(-|m|\lambda_-)$.
For $E>2$, the exponent $\lambda_+$ has an imaginary part equal to $i\pi$. Consequently, the wavefunction decreases exponentially {\em and} oscillates with $m$: $\psi_m\propto (-)^m\exp(-|m|\text{Re}\lambda_+)$.
\subsubsection{Three on-site defects}
\label{sec:three_defects}
Here, we consider the case of three attractive impurities at $m=0$ and $m=\pm 1$ with energies $U_0<0$ and $U_1<0$, correspondingly (see Fig.~\ref{three_on-site}a).  We use Ansatz (\ref{Ansatz}) to write down a closed system of equations for the three defective sites
\bse
\bea
E \psi_0 &= U_0 \psi_0 -  \psi_{1}-\psi_{-1} ,\label{TB_3a} \\
E \psi_{\pm 1} &= U_1 \psi_{\pm 1} - \psi_0 - \psi_{\pm 1}e^{-\lambda}, \label{TB_3b}
\eea
\ese
complemented again by Eq.~(\ref{reg}). Since the impurities are attractive, the bound states have energies in the interval $E<-2$. Correspondingly, we choose $\lambda_{+}$ for the exponent of the wave function in Eq.~(\ref{el}).

\begin{figure}[htbp]
\centering
\includegraphics[scale=0.4]{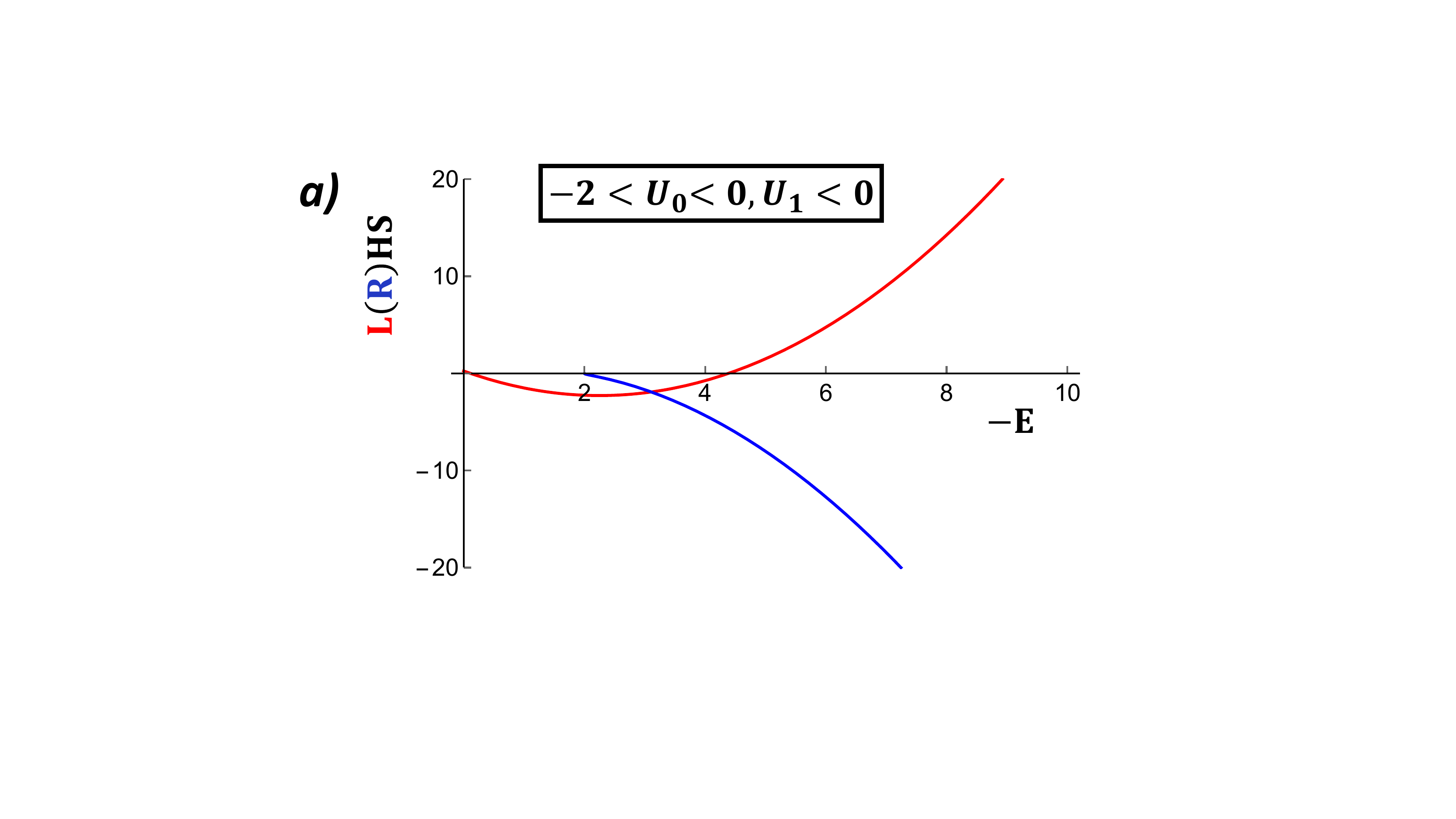}
\includegraphics[scale=0.4]{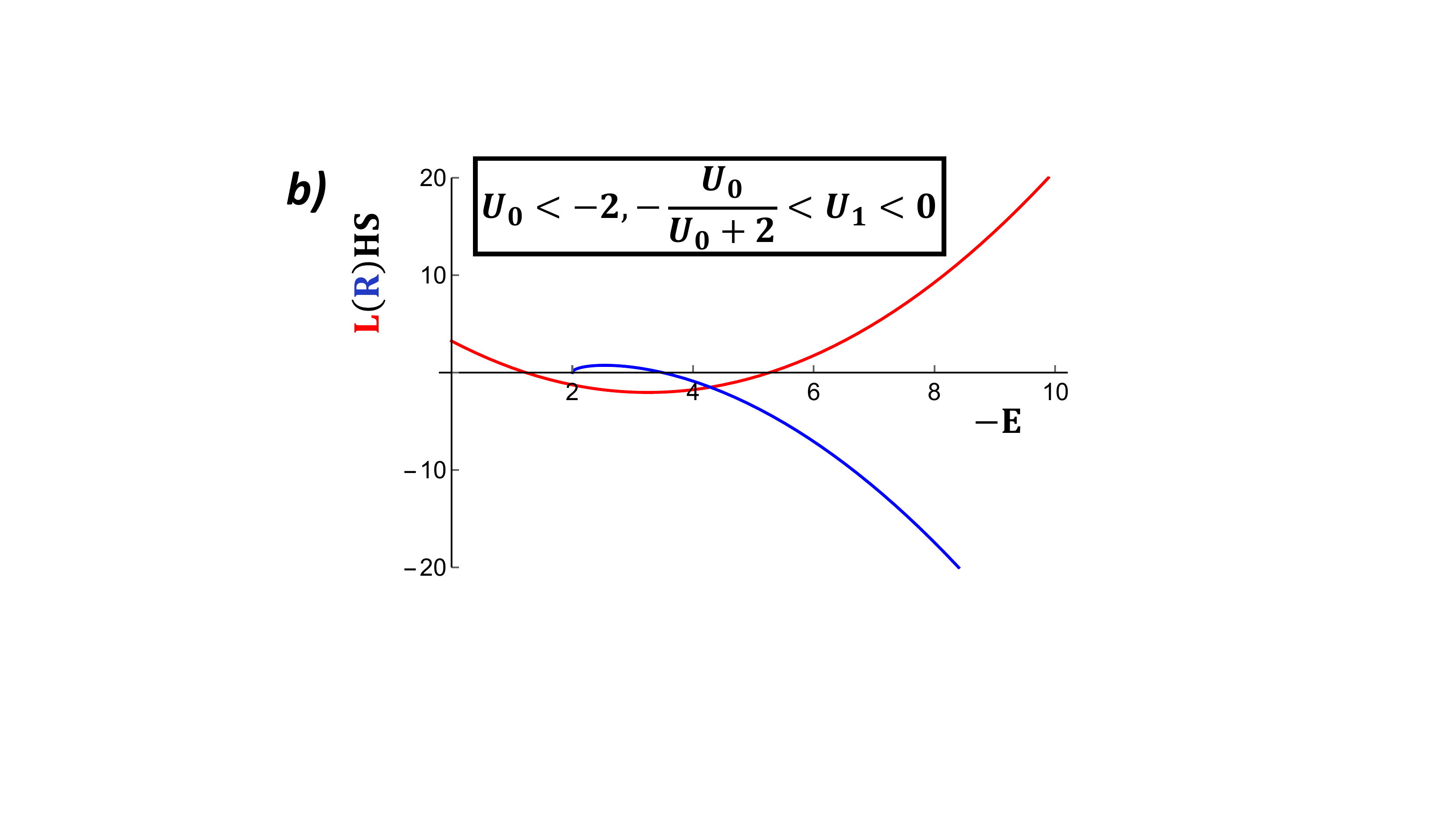}
\includegraphics[scale=0.4]{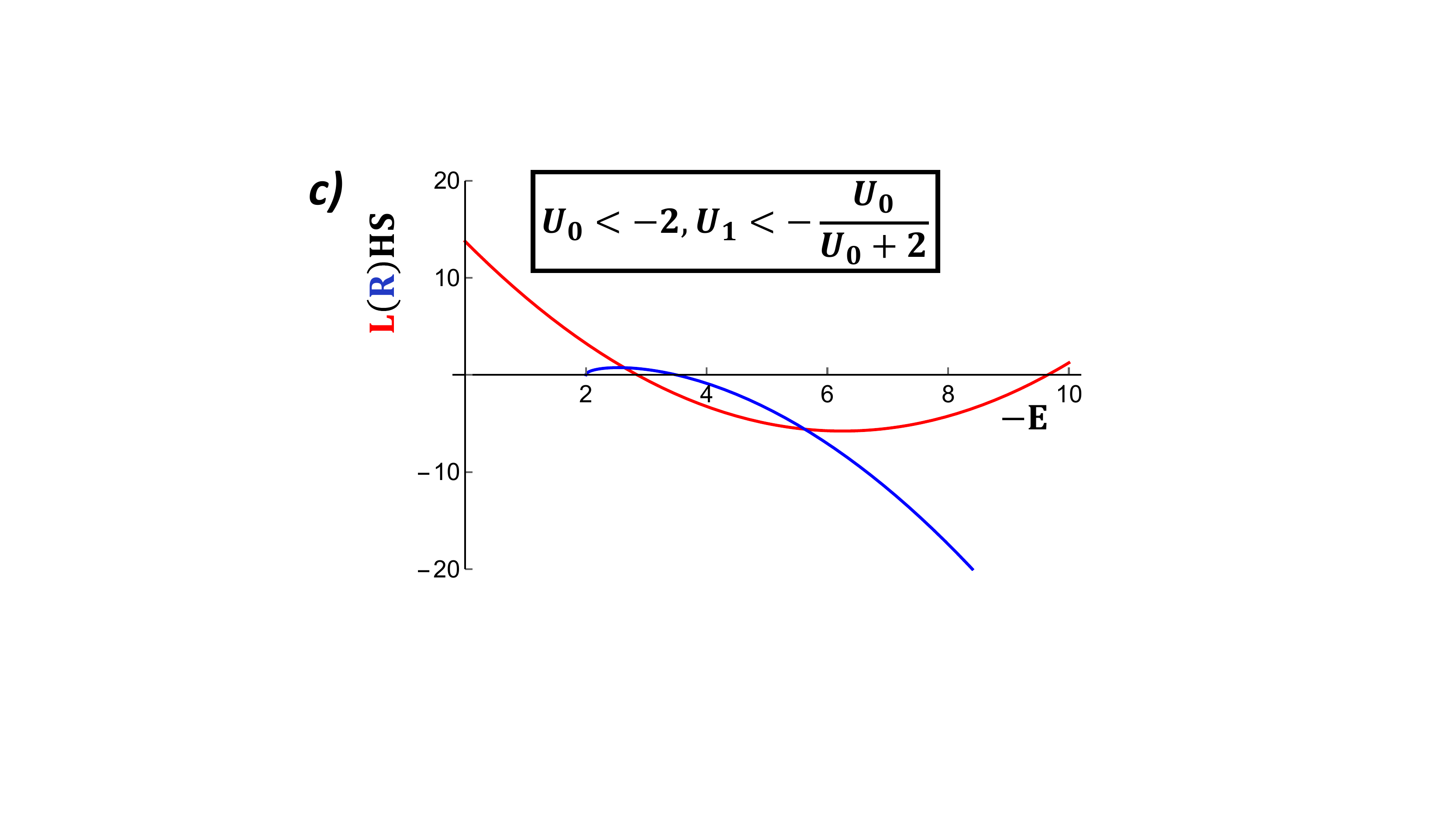}
\caption{\label{only_three_on-site} Graphic solution of Eq.~(\ref{yy}).}
\end{figure}

For even-parity solutions, $\psi_1=\psi_{-1}$. Substituting this relation
into Eqs.~(\ref{TB_3a}) and (\ref{TB_3b}) and using Eq.~(\ref{el}), we obtain an equation for the corresponding eigenenergies \begin{equation}
\label{yy}
\frac{E^{2}}{2} -\bigg(\frac{U_0}{2}+U_1\bigg) E+U_0 U_1 - 2 = (E-U_0) \sqrt{\frac{E^{2}}{4} - 1}.
\end{equation}
The graphic solution of Eq.~(\ref{yy}) is shown in Fig.~\ref{only_three_on-site}. The bound states are below the band edge, which means that $E<-2$.  Consequently, the right-hand side (RHS) of Eq.~(\ref{yy}) is negative-definite for $-2<U_0<0$, as shown in Fig.~\ref{only_three_on-site}a. The left-hand side (LHS) is a parabola which is equal to $P=U_1(U_0+2)+U_0$ at $E=-2$. For $U_1<0$ and $-2<U_0<0$, $P$ is negative, and therefore Eq.~(\ref{yy}) has only one root. This is the lowest-energy bound state, which occurs even for an infinitesimally small $U_0$. If $U_0<-2$, the RHS  of Eq.~(\ref{yy}) vanishes at $E=-2$ and $E=U_0$, and has a maximum in between these two points. For $-U_0/(U_0+2)<U_1<0$, $P$ is still negative and thus the equation has only one root, see Fig.~\ref{three_on-site}b. For $U_1<-U_0/(U_0+2)$, $P$ is positive and the equation has two roots, see  Fig.~\ref{three_on-site}c.

For odd-parity solutions, $\psi_1=-\psi_{-1}$.  Substituting this relation into Eq.~(\ref{TB_3a}), we obtain $E\psi_0=U_0\psi_0$, which has two solutions: $E=U_0$ and $\psi_0=0$. However,  substituting $\psi_1=-\psi_{-1}$ into Eq.~(\ref{TB_3b}), we find that equations for $\psi_{\pm}$ are compatible only if $\psi_0=0$. Therefore, $\psi_0$ for an odd-parity solution and each of the two equations in Eq.~(\ref{TB_3b}) combined with Eq.~(\ref{el}) yields 
\beq
E = U_1 +1/U_1,
\eeq 
which corresponds to a bound state if $U_1<-1$. This is one of the two ``extra'' bound states which occur only if impurities are strong enough.

Combining all these cases together and restoring $J$, we obtain the phase diagram shown in Fig.~\ref{phase-diagram}.

\subsection{Bond defects only}
\label{sec:bondsA}
Here, we consider a model with two defective bonds connecting  the $m=0$ site with two adjacent sites at $m=\pm 1$ (see Fig.~\ref{three_on-site}b).
The Schroedinger equations read
\bea
\label{bond_disorder}
E \psi_0 &=& - J' (\psi_{-1} + \psi_{1}),\nn \\
E \psi_{\pm 1} &=& - J' \psi_0 - \psi_{\pm 2},\nn \\
E \psi_{\pm(|m|+2)} &=& -(\psi_{\pm(|m|+1)} + \psi_{\pm(|m|+3)}),
\eea
where $J'>0$ is the amplitude of hopping between the $m=0$ and $m=\pm 1$ sites, and the hopping amplitude for normal bonds ($J$) is set to unity.   
A simple check shows that a non-trivial solution is possible only for an even-parity state with $\psi_1=\psi_{-1}$, in which case we can set $\psi_0=1$. This yields 
\begin{equation}
E = \pm \frac{2J'^{2}}{\sqrt{2J'^{2} - 1}}.
\end{equation}
The energies of the bound states are outside the band only if $|J'|>1$, i.e., only if the defect bonds are stronger than the normal ones.

\subsection{Both on-site and bond defects}
\label{comp}
\subsubsection{Single on-site and two bond defects}
\label{sec:1i2b}
To understand the competition between on-site and bond defects, we consider here a model with a single on-site defect connected by defective bonds to the rest of the lattice, as depicted in Fig.~\ref{three_on-site}c. The Schroedinger equations in this case read
\bse
\bea
E \psi_0 &= & U_0 \psi_0 - J' (\psi_{-1} + \psi_{1}),\label{competition_a} \\
E \psi_{\pm 1} &=& - J' \psi_0 - \psi_{\pm 2},\label{competition_b}\\
E \psi_{\pm(|m|+2)} &=& -(\psi_{\pm(|m|+1)} + \psi_{\pm(|m|+3)}),\label{competition_c}
\eea
\ese
where again we set $J=1$.
 It can be readily seen that Eqs.~(\ref{competition_a}-\ref{competition_c}) have no odd-parity solutions. Indeed, such a solution corresponds to $\psi_0=0$ which, upon substituting into Eq.~(\ref{competition_b}) and using $\psi_{\pm 2}=e^{-\lambda}\psi_{\pm 1}$, yields $E+e^{-\lambda}=0$.  However, the last equation is not compatible with Eq.~(\ref{reg}). We thus focus on the even-parity solution with $\psi_1=\psi_{-1}$ and set $\psi_0=1$. We are  interested in bound states below the conduction band, whose wavefunction decays exponentially with exponent $\lambda_-$ defined by Eq.~(\ref{el}).

Eliminating $\psi_1$ from Eqs.~(\ref{competition_a}) and (\ref{competition_b}), we obtain an eigenvalue equation 
\begin{equation}
\label{intermediate}
  E(1-J'^2)-U_0 = 2J'^2 \sqrt{\frac{E^2}{4} - 1}.
\end{equation}
There is a negative-$E$ solution if $J'^2>1$ and $U_0<0$. Indeed,  the RHS of Eq.~(\ref{intermediate}) vanishes at $E=-2$, whereas the LHS is positive. On the other hand, the RHS behaves as $-EJ'^2$ for $-E\to\infty$, while the LHS is always below this value. Therefore, an intersection of the two curves must have happened at finite $-E$.  For $J'^2<1$, a solution exists only if $U_0< - 2(1-J'^2) $. If $U_0 <-2$, this condition is always satisfied but, if $-2<U_0 < 0$, the bound state disappears if  $J'^2 < 1 + U_0/2$.
To conclude,  there is always a bound state below the conduction band for any $J'$, if the impurity is sufficiently strong, i.e.,   $U_0<-2$. However, if the impurity is sufficiently weak, i.e., $-2<U_0<0$, there is no bound state for $J'^2 < 1 + U_0/2$.

\subsubsection{Three on-site and two bond defects}
\label{com}
\begin{figure}[htbp]
\vspace{0.2in}
\centering
\includegraphics[scale=0.35]{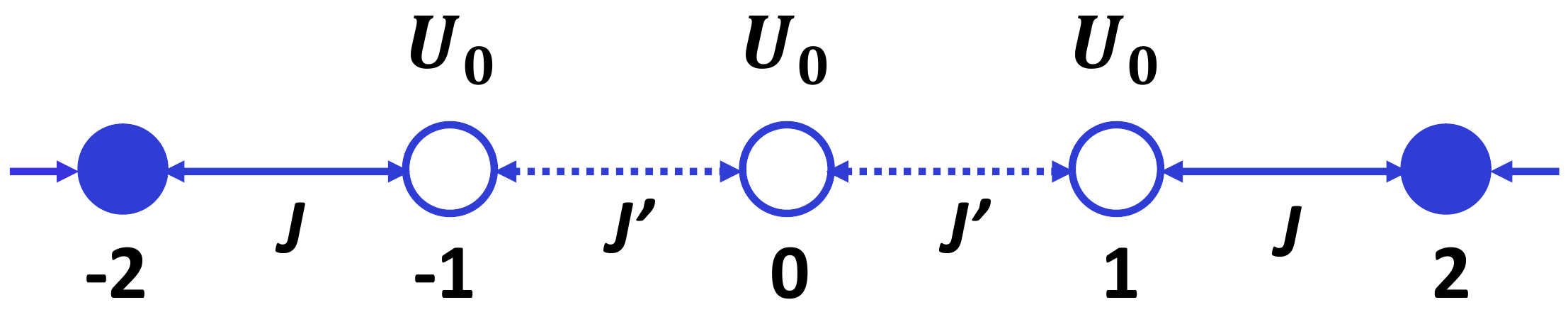}
\caption{\label{three_on-site_bond} One-dimensional tight-binding model with three identical impurities connected by defective bonds.  All the other bonds are normal, i.e., $J=1$.}
\end{figure}

Here, we consider a slightly more complicated model with three identical and attractive impurities connected by defective bonds, as shown in Fig.~\ref{three_on-site_bond}. This model illustrates a transition in the number of bound states,  from three to two and eventually to one.

The system is described by the following set of equations:
\begin{equation}
\label{compete}
\begin{split}
E \psi_0 &= U_0 \psi_0 - J'(\psi_{-1} + \psi_{1}), \\
E \psi_{\pm 1} &= U_0 \psi_{\pm 1} + J' \psi_0 - \psi_{\pm 2}, \\
E \psi_{\pm(|m|+2)} &= -(\psi_{\pm(|m|+1)} + \psi_{\pm(|m|+3)}),
\end{split}
\end{equation}
complemented by Eq.~(\ref{ansatz}) and Eq.~(\ref{el}) for $\exp(-\lambda_-)$. 
The eigenvalues of even- and odd parity states satisfy
\bse
\bea
(E -U_0)(E - U_0 + e^{-\lambda_-}) - 2J'^2 &=& 0,\label{m2a}\\
E -U_0 + e^{-\lambda_-} &=& 0\label{m2b},
\eea
\ese
correspondingly.
Equation (\ref{m2b}) yields $E = U_0 +1 / U_0$, which satisfies the condition for the bound state to be below the band edge,  i.e., $E<-2$, for $U_0<-1$. Since the odd-parity eigenvalue does not depend on $J'$, it is not affected by the competition between on-site and bond defects. 
Equation (\ref{m2a}) for the even-parity eigenvalue is reduced to
\begin{equation}
\label{ee}
\frac{E^2}{2} - \frac{3}{2} EU_0+U_0^2  - 2J'^2 = ( E-U_0 ) \sqrt{\frac{E^2}{4} - 1}.
\end{equation}
To solve  this equation, we use the same argument as given in  Sec.~\ref{sec:three_defects} for case of three attractive impurities. The RHS is negative- definite for $E<-2$ and $-2<U_0<0$ and vanishes at $E=-2$. The LHS is an upward parabola which takes a value  of $\tilde P=U_0^2+3U_0+2-2J^2$ at $E=-2$.  For $J'^2>J^2_c\equiv (U_0^2+3U_0+2$)/2, $\tilde P$ is negative and hence Eq.~(\ref{ee}) has only one root. For $U_0<-2$, the RHS of Eq.~(\ref{ee}) vanishes at $E=-2$ and $E=U_0$, and has a maximum in between these two points. For $J'^2>J_c^2$, $\tilde P$ is still negative and thus the equation still has only one root. If $U_0<-2$ and $J^2<J_c^2$, $\tilde P$ is positive and  the equation has two roots. Combining these arguments with the odd-parity case, we conclude that for $-1<U_0<0$ and $J^2>J_c^2$, the system has only one bound state.  For $U_0<-1$ and $J'^2>J_c^2$, the number of the bound states increases to two. Finally,  there are three bound states for $U_0<-1$ and $J'^2<J_c^2$.

\section{Explicit expressions for the frequencies of chiral spin modes in the $s$-wave approximation}
\label{app:s_wave}
In this Appendix, we present explicit expressions for the frequencies of chiral spin modes within the $s$-wave approximation for the Landau function [Eq.~(\ref{swave})]. Equating the determinant of the 3$\times$3 matrix in Eq.~(\ref{s-wave matrix equation}) to zero, we find
\begin{widetext}
\bea
\label{collective_modes_swave}
\tilde \Omega_0^2 &=& P - \frac{1}{6\times 2^{2/3}} \frac{1}{F_0^a} \Bigg[ \frac{Q}{ \big(\text{Re} \left\{Z\right\} \big)^2 + \big(\text{Im} \left\{Z\right\} \big)^2 } - 2^{-2/3} \Bigg] \Big[ \text{Re} \left\{Z\right\} + \sqrt{3} {\,} \text{Im} \left\{Z\right\} \Big],\nn\\
\tilde \Omega_{+1}^2 &=& \bigg[ 1 + \frac{F_0^a}{2} \bigg]\Delta_R^{2} + \bigg[ 1+ \frac{2}{F_0^a} \bigg] \Delta_Z^{*2}, \nn\\
\tilde \Omega_{-1}^2 &=& P + \frac{1}{3\times 2^{2/3}} \frac{1}{F_0^a} \Bigg[ \frac{Q}{ \big(\text{Re} \left\{Z\right\} \big)^2 + \big(\text{Im} \left\{Z\right\} \big)^2 } - 2^{-2/3} \Bigg] \text{Re} \left\{Z\right\},\label{B1}
\eea
\end{widetext}
where $Z = \Big( R + \sqrt{4 Q^3 + R^2} \Big)^{1/3}$ and
\begin{widetext}
\bea
P &=& \frac{1}{6} \Big[ 6 + 5 F_0^a \Big] \Delta_R^{2} -  \frac{1}{3 F_0^a} \Big[ 2 + 5 F_0^a + \big(F_0^a\big)^2 - \big(F_0^a\big)^3 \Big] \Delta_Z^{*2},\nn \\
Q &=& -\big(F_0^a\big)^4 \Delta_R^{4} - 4 \big(F_0^a\big)^2 \Big[ 2 + 8 F_0^a + 7 \big(F_0^a\big)^2 + 2 \big(F_0^a\big)^3 \Big] \Delta_R^{2}\Delta_Z^{*2} - 4 \Big[ 4 + 32 F_0^a + 68 \big(F_0^a\big)^2 + 60 \big(F_0^a\big)^3 + 27 \big(F_0^a\big)^4, \nn\\
&&+ 7 \big(F_0^a\big)^5 + \big(F_0^a\big)^6 \Big] \Delta_Z^{*4},\nn \\
R &=& 2 \big(F_0^a\big)^6 \Delta_R^{6} + 12 \big(F_0^a\big)^4 \Big[ 2 + 8 F_0^a + 7 \big(F_0^a\big)^2 + 2 \big(F_0^a\big)^3 \Big] \Delta_R^{4}\Delta_Z^{*2} - 12 \big(F_0^a\big)^2 \Big[ 136 + 512 F_0^a + 776 \big(F_0^a\big)^2 + 612 \big(F_0^a\big)^3,\nn \\
&&+ 261 \big(F_0^a\big)^4 + 55 \big(F_0^a\big)^5 + 4 \big(F_0^a\big)^6 \Big] \Delta_R^{2}\Delta_Z^{*4} + 8 \Big[ 16 + 192 F_0^a + 792 \big(F_0^a\big)^2 + 1480 \big(F_0^a\big)^3 + 1392 \big(F_0^a\big)^4 + 630 \big(F_0^a\big)^5\nn \\
&&+ 62 \big(F_0^a\big)^6 - 57 \big(F_0^a\big)^7 - 21 \big(F_0^a\big)^8 - 2 \big(F_0^a\big)^9 \Big] \Delta_Z^{*6}.
\eea
\end{widetext}
The modes are labeled according to Eq.~(\ref{freqR}) for the case of zero magnetic field. In the limit of $B=0$, $\tilde \Omega_0$ approaches $\Omega_0$ in Eq.~(\ref{om0s}), while $\tilde \Omega_{\pm 1}$ become degenerate and equal to $\Omega_{\pm 1}$ in Eq.~(\ref{om1s}). The mode frequencies in Eq.~(\ref{B1}) are plotted in Fig.~\ref{R+B_s_wave} as a function of the magnetic field. 

\section{Fermi-liquid interaction in the $m=0$ and $m=1$ channels}
\label{app:s_p_wave}
In this Appendix, we provide the details of deriving the equations of motion for the case of the Landau function in the $s+p$-wave approximation, as specified by Eq.~(\ref{sp}).  
For simplicity, we consider a situation when only the in-plane magnetic field and Rashba SOC are present but Dresselhaus  SOC is absent. Applying Eq.~(\ref{sp}) to Eq.~(\ref{tight binding}), we obtain
\begin{widetext}
\begin{equation}
\begin{split}
&-\Omega^2 \psi_0 = -\bigg[\Delta_R^{*2} \big(1+ F_1^a \big) \big(1+F_0^a \big) + \Delta_Z^{*2} \Big(1+ F_0^a \big(2+F_0^a \big) \Big)\bigg] \psi_0 + \Delta_R^* \Delta_Z^* \bigg[1 + \frac{1}{2} F_1^a \big(3+F_1^a \big) + \frac{1}{2} F_0^a \big(1+F_1^a \big) \bigg] \Big(\psi_{-1} + \psi_{1} \Big), \\
&-\Omega^2 \psi_{\pm 1}= -\bigg[\Delta_R^{*2} \big(1+ F_1^a \big) \big(1+\frac{1}{2}F_0^a \big) + \Delta_Z^{*2} \Big(1+ F_1^a \big(2+F_1^a \big) \Big) \bigg] \psi_{\pm 1} + \Delta_R^* \Delta_Z^* \bigg[1 + \frac{1}{2} F_1^a \big(1+F_0^a \big) + \frac{1}{2} F_0^a \big(3+F_0^a \big) \bigg] \psi_{0} \\
&{\,\,\,\,\,\,\,\,\,\,\,\,\,\,\,\,\,\,\,\,\,\,\,\,\,\,\,\,\,\,\,}+ \Delta_R^* \Delta_Z^* \bigg[1+\frac{1}{2} F_1^a \bigg]\psi_{\pm 2}, \\
&-\Omega^2 \psi_{\pm 2} = -\bigg[\Delta_R^{*2} \bigg(1+ \frac{1}{2}F_1^a \bigg) + \Delta_Z^{*2} \bigg] \psi_{\pm 2} + \Delta_R^* \Delta_Z^* \bigg[1+\frac{1}{2} F_1^a \big(3+F_1^a \big) \bigg] \psi_{\pm 1} + \Delta_R^* \Delta_Z^* {\,} \psi_{\pm 3}, \\
&-\Omega^2 \psi_{\pm (|m|+3)} = -\Big[\Delta_R^{*2} + \Delta_Z^{*2} \Big] \psi_{\pm (|m|+3)} + \Delta_R^* \Delta_Z^* \Big(\psi_{\pm (|m|+2)} + \psi_{\pm (|m|+4)} \Big).
\end{split}
\end{equation}
\end{widetext}
To close the system, we assume that the wavefunctions of bound states fall off with distance exponentially: $\psi_{\pm (|m|+3)} = e^{-(|m|+1)\lambda}\psi_{\pm 2}$ for $m \geq 0$,  where $e^{-\lambda}$ is given by Eq.~(\ref{lambda}). To reduce the system size from 5$\times$5 to 3$\times$3, we eliminate $\psi_2$ in favor of $\psi_1$, i.e.,  we set $\psi_{\pm 2} = 
\gamma\psi_{\pm 1}$, where
\begin{equation}
\gamma= \frac{\Delta_R^* \Delta_Z^* \Big(1+\frac{1}{2}F_1^a \big(3+F_1^a \big) \Big)}{\big(\Delta_R^{*2} + \Delta_Z^{*2} - \Omega^2 \big) - \Delta_R^* \Delta_Z^* e^{-\lambda} + \frac{1}{2}\Delta_R^{*2}F_1^a}.
\end{equation}
The frequencies of the collective modes are given by zeroes of the determinant of the resulting 3$\times$3 system:
\begin{widetext}
\begin{equation}
\begin{split}
&\text{Det}\left[\begin{array}{ccc}
G_{11} & G_{12} & 0 \\
G_{21} & G_{22} & G_{23} \\
0 & G_{32} & G_{33}
\end{array} 
\right]=0
 \\
\text{where}\; G_{11} &= G_{33} = \Omega^2 - \bigg[\Delta_R^{*2} \big(1+ F_1^a \big) \big(1+\frac{1}{2}F_0^a \big) + \Delta_Z^{*2} \Big(1+ F_1^a \big(2+F_1^a \big) \Big) \bigg] + \Lambda, \\
G_{12} &= G_{32} = \Delta_R^* \Delta_Z^* \bigg[1 + \frac{1}{2} F_1^a \big(1+F_0^a \big) + \frac{1}{2} F_0^a \big(3+F_0^a \big) \bigg], \\
G_{21} &= G_{23} = \Delta_R^* \Delta_Z^* \bigg[1 + \frac{1}{2} F_1^a \big(3+F_1^a \big) + \frac{1}{2} F_0^a \big(1+F_1^a \big) \bigg], \\
G_{22} &= \Omega^2 - \bigg[ \Delta_R^{*2} \big(1+ F_1^a \big) \big(1+F_0^a \big) + \Delta_Z^{*2} \Big(1+ F_0^a \big(2+F_0^a \big) \Big)\bigg], \\
\text{and}\; \Lambda &= \frac{2 \Delta_R^{*2}\Delta_Z^{*2} (1+\frac{1}{2} F_1^a) (1+\frac{1}{2} F_1^a (3+F_1^a))}{(\Delta_R^{*2} + \Delta_Z^{*2} - \Omega^2) + \Delta_R^{*2} F_1^a + \sqrt{(\Delta_R^{*2} + \Delta_Z^{*2} - \Omega^2)^2 - 4 \Delta_R^{*2}\Delta_Z^{*2}}}.
\end{split}
\end{equation}
\end{widetext}
\bibliography{ref}


\end{document}